%% file: header.tex
\documentclass[acmtog]{acmart}
\acmSubmissionID{715}

\usepackage{standalone}
\usepackage{tikz}
\usepackage{graphicx}
\usepackage{subcaption}
\usepackage{amsmath,bm}
\usepackage{booktabs} 
\usepackage{eucal}
\usepackage{soul}
\usepackage{xcolor}

\usepackage{wrapfig}
\usepackage[switch]{lineno}
\usepackage[boxed,ruled,linesnumbered]{algorithm2e} 

\SetCommentSty{mycommfont}
\let\oldnl\nl

\newcommand{\nonl}{\renewcommand{\nl}{\let\nl\oldnl}}
\newlength\savedwidth
\newcommand\whline[1]{\noalign{\global\savedwidth\arrayrulewidth
                               \global\arrayrulewidth #1} %
                      \hline
                      \noalign{\global\arrayrulewidth\savedwidth}}
                      
\usepackage[ruled]{algorithm2e} 

\SetAlFnt{\small}
\SetAlCapFnt{\small}
\SetAlCapNameFnt{\small}
\SetAlCapHSkip{0pt}

\acmJournal{TOG}

\begin{document}
\title{JGS2: Near Second-order Converging Jacobi/Gauss-Seidel for GPU Elastodynamics}

\author{Lei Lan}
\orcid{0009-0002-7626-7580}
\affiliation{%
 \institution{University of Utah}
 \country{USA}}

\author{Zixuan Lu}
\orcid{0000-0003-0067-0242}
\affiliation{%
 \institution{University of Utah}
 \country{USA}}
\email{birdpeople1984@gmail.com}

\author{Chun Yuan}
\orcid{0009-0009-1134-0442}
\affiliation{%
 \institution{ University of Utah}
 \country{USA}}
\email{yuanchunisme@gmail.com}

\author{Weiwei Xu}
\orcid{0000-0003-3756-3539}
\affiliation{%
 \institution{State Key Lab of CAD\&CG, Zhejiang University}
 \country{China}}
\email{xww@cad.zju.edu.cn}

\author{Hao Su}
\orcid{0000-0002-1796-2682}
\affiliation{%
 \institution{UCSD}
 \country{USA}}
\email{haosu@eng.ucsd.edu}

\author{Huamin Wang}
\orcid{0000-0002-8153-2337}
\affiliation{%
 \institution{Style3D Research}
 \country{China}}
\email{wanghmin@gmail.com}

\author{Chenfanfu Jiang}
\orcid{0000-0003-3506-0583}
\affiliation{%
 \institution{UCLA}
 \country{USA}}
\email{chenfanfu.jiang@gmail.com}

\author{Yin Yang}
\orcid{0000-0001-7645-5931}
\affiliation{%
 \institution{University of Utah}
 \country{USA}}
\email{yangzzzy@gmail.com}

\begin{abstract}
In parallel simulation, convergence and parallelism are often seen as inherently conflicting objectives. Improved parallelism typically entails lighter local computation and weaker coupling, which unavoidably slow the global convergence. This paper presents a novel GPU algorithm that achieves convergence rates comparable to fullspace Newton's method while maintaining good parallelizability just like the Jacobi method. Our approach is built on a key insight into the phenomenon of \emph{overshoot}. Overshoot occurs when a local solver aggressively minimizes its local energy without accounting for the global context, resulting in a local update that undermines global convergence. To address this, we derive a theoretically second-order optimal solution to mitigate overshoot. Furthermore, we adapt this solution into a pre-computable form. Leveraging Cubature sampling, our runtime cost is only marginally higher than the Jacobi method, yet our algorithm converges nearly quadratically as Newton's method. We also introduce a novel full-coordinate formulation for more efficient pre-computation. Our method integrates seamlessly with the incremental potential contact method and achieves second-order convergence for both stiff and soft materials. Experimental results demonstrate that our approach delivers high-quality simulations and outperforms state-of-the-art GPU methods with {$\bm{50\times}$ to $\bm{100\times}$} better convergence.
\end{abstract}

%
%

\ccsdesc[500]{Computing methodologies~Physical simulation}

%
%
\begin{teaserfigure}
\centering
\includegraphics[width=\textwidth]{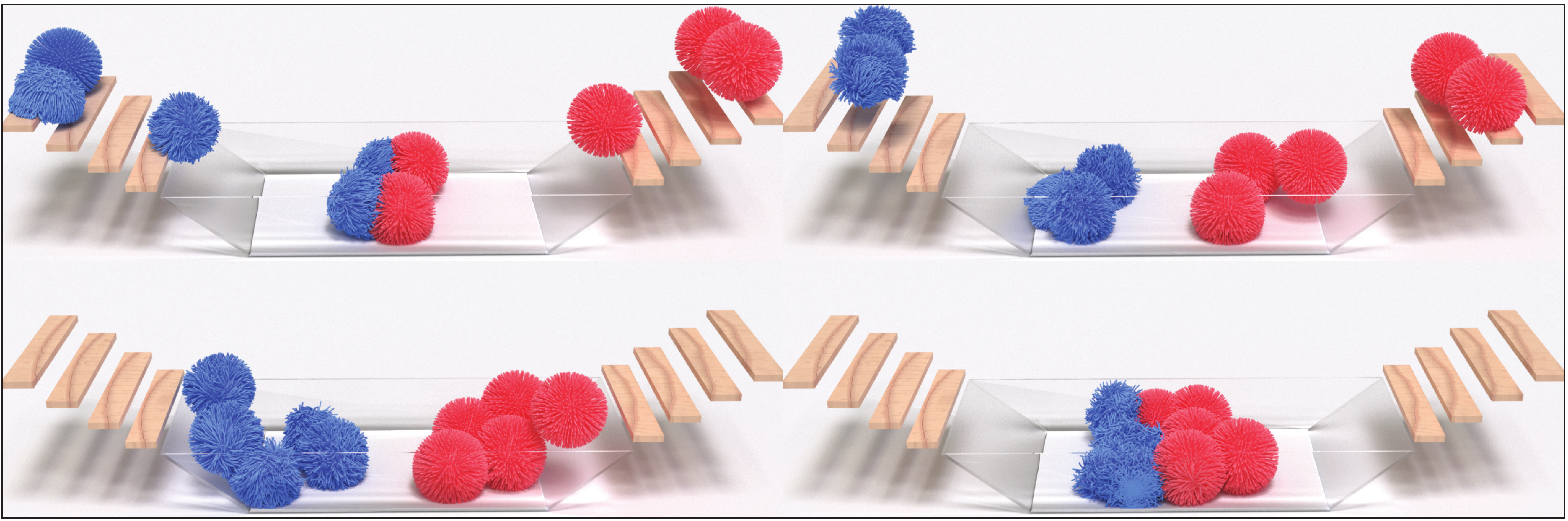}
\captionof{figure}{\textbf{Soft and stiff puffer balls.}~~This paper presents a novel GPU-based parallel algorithm for elastic body simulation. We are inspired by a numerical issue of overshoot, which is the major reason behind the slow convergence of parallel solvers. Overshoot refers to the situation where local relaxation becomes over-aggressive --- the reduction of local energy gets outweighed by the energy increase at other regions on the deformable object. We offer a second-order optimal solution to resolve this issue so that a parallel iteration becomes as convergent as a global Newton solve. Based on this observation, we carefully re-design the computation procedure, making this solution efficient and pre-computable. As a result, our method possesses superior parallelism (as using the Jacobi method) and near second-order convergence (as using global Newton's method). It constantly converges $50\times$ to $100\times$ faster than the state-of-the-art GPU methods, and our advantage is more significant for stiff simulations. The teaser figure shows a representative example. In this experiment, 10 puffer balls slide down into a glass tank. There are $3.5$M tetrahedron elements in this example, and the time step size is $1/120$. Blue balls are 20 times softer than red balls. Vertex block descent fails to converge at this time step. If all the puffer balls are soft ones, our method is $122\times$ faster than vertex block descent.}
\label{fig:teaser}
\end{teaserfigure}

\keywords{GPU simulation, Second-order Jacobi method, Newton's method, Numerical optimization, Parallel computation}

\maketitle
\input{introduction}

\input{related}
\input{overundershoot}

\input{second}
\input{cubature}

\input{pre_computation}

\input{collision}
\input{exp}
\input{conclusion}

\bibliographystyle{ACM-Reference-Format}
\bibliography{ref}

\end{document}

%% file: introduction.tex
\section{Introduction}\label{sec:intro}
Since its inception, physics-based simulation has been synonymous with high computational cost. Integrating such techniques into time-critical applications stands a significant challenge. Since then, various techniques aimed at improving simulation performance have been developed. For instance, it is possible to simplify the nonlinearity of the material to reuse a pre-factorized system matrix such as stiffness warping~\cite{muller2002stable,choi2005modal}, or we can build a reduced-order model using much fewer simulation degrees of freedom (DOFs)~\cite{pan2015subspace,barbivc2005real,an2008optimizing}. Despite achieving orders-of-magnitude speedups, these methods often come at the cost of compromised accuracy to some extent. In other words, they trade physical precision for performance gain. The advent of GPGPU has brought new opportunities to the field of simulation. Equipped with a large number of processing units, GPUs excel in handling massive sub computing tasks simultaneously. One-stop solvers like Newton's method using direct Hessian factorization do not fit this new computation paradigm, and nearly all the GPU algorithms opt for iterative and parallel numerical procedures. Two representative examples are the Jacobi and Gauss-Seidel (GS) methods. 

Here, we consider Jacobi or GS as nonlinear relaxation schemes, where the target function is divided into small sub-problems with shared DOFs. When put into the context of quadratic optimization, they become iterative linear solvers~\cite{greenbaum1997iterative}. Jacobi scheme solves each sub-problem independently. Each shared DOF has several local replicas at sub-problems, which are averaged by the end of the iteration. The classic GS routine, on the other hand, solves sub-problems sequentially --- the newly updated DOF values participate in the following local solves. Parallel GS leverages graph coloring algorithms that group sub-problems without DOFs sharing so that GS updates within the group can be executed in parallel.

The key ingredient of an effective GPU simulation algorithm is always a wise trade-off between parallelization and convergence. Strategies like increasing the size of local sub-problems~\cite{lan2023second} or making sub-problems more overlapping~\cite{luo2017physics} favor convergence, helping the algorithm become more effective for stiff instances. Downsizing sub-problems~\cite{chen2024vertex} or delaying the information exchange~\cite{fei2021principles} enhances the parallelization at the cost of more iterations or even divergence. It is believed that one can not achieve the best parallelization and convergence at the same time. 

In this paper, we show a novel GPU algorithm that substantially narrows, if not closes, the gap between convergence and parallelization. Our key observation is local \emph{overshoot} i.e., fully solving local sub-problems without knowing the global information. Overshoots negatively impact global convergence and stand as the main culprit for the slow convergence of most existing GPU algorithms. However, this issue has been overlooked and went largely unnoticed. We propose a solution that makes local computation globally aware by pre-building a reduced model for each local sub-problem. This strategy is material-aware and behaves equally well for both extremely stiff and soft problems. When the optimization problem can be well-approximated by a quadratic form, i.e., it falls within the scope of Newton's method, our approach achieves near-optimal convergence. As a result, our method converges at or near the rate of the full Newton's method while being as parallelizable as Jacobi or GS. We have tested our method in various simulation scenes. The experiment results reported are encouraging  --- our method converges $50\times$ to $100\times$ faster than the state-of-the-art GPU algorithms, paving the path to real-time and high-resolution simulation without accuracy compromises. We demonstrate the efficiency and efficacy of our algorithm in the context of elastodynamic simulation using finite element method (FEM)~\cite{bathe2006finite} however, the proposed method is readily applicable in other simulations problems such as cloth/thin shell simulation, rod simulation, MPM (material point method), and fluid simulation.

%% file: related.tex
\section{Related Work}\label{sec:related}
High-resolution deformable bodies house a large number of two-way coupled unknown DOFs, and implicit time integration methods like backward Euler~\cite{baraff1998large} or Newmark~\cite{hughes2012finite} are commonly used for improved numerical stability. This results in a global (often sparse) nonlinear system. Solving this system at each time step becomes the major bottleneck of the simulation pipeline. 

An effective strategy is to avoid a full linear solve in classic Newton's method. Following this idea, \citet{hecht2012updated} proposed a lagged factorization scheme that reuses existing Cholesky factorization to save the computation. \citet{chen2024trust} exploited the global quadratic approximation quality to control the local eigenvalue projection in projected Newton. Multi-resolution~\cite{capell2002multiresolution,grinspun2002charms} and multigrid solvers project fine-grid residual errors onto a coarser grid, on which linear or nonlinear iterations are more effective~\cite{zhu2010efficient,xian2019scalable,tamstorf2015smoothed,wang2020hierarchical,bolz2003sparse}. Quasi-Newton methods use Hessian approximates, instead of the exact Hessian, to estimate a good search direction~\cite{liu2017quasi,Li2019DOT,wang2020hierarchical}. \citet{zhang2024progressive} took the fine-level energy into account when computing the coarse-level update in the multigrid solver for cloth and thin-shell simulation. Those methods are mostly CPU-based and seek performance gain through trading the numerical accuracy. As many graphics applications emphasize more on visual plausibility, such a trade-off is reasonable and practical.  

Simplifications of the underlying elasticity model also lead to many important simulation techniques. A classic example is stiffness warping~\cite{muller2002stable}, which can be viewed as a simplified co-rotated material. It allows the re-use of the rest-shape stiffness matrix for rotational deformation. Stiffness warping can also be combined with modal analysis to enable real-time simulations~\cite{choi2005modal}. \citet{chao2010simple} designed a simplified material model measuring the distance of linear deformation and rotation. This concept is similar to the shape matching algorithm~\cite{muller2005meshless}, where the deformation energy is defined based on the nearest rigid body transformation. PBD (position-based dynamics)~\cite{muller2007position} and extended PBD (XPBD)~\cite{macklin2016xpbd} regard the elastic energy as a set of \emph{compliant constraints} and use the steepest descent or gradient descent to update the vertex positions at each constraint. This method is later generalized for other simulation problems, including fluid~\cite{macklin2013position}, rigid bodies~\cite{muller2020detailed}, and MPM~\cite{yu2024xpbi}. Similarly, projective dynamics (PD) treats the elasticity energy as a collection of quadratic constraints. This assumption allows the user to separate the constraint projection and the distance measure into local and global steps~\cite{bouaziz2014projective}. The key benefit of PBD and PD is the decoupling of DOFs in different constraints. As a result, both methods can be parallelized on the GPU~\cite{fratarcangeli2018parallel,fratarcangeli2016vivace,wang2015chebyshev}. In other words, the idealization of the underlying material model eases the solving procedure. However, the generalization of those methods to more complicated and real-world materials is less intuitive, and a careful re-formulation is needed~\cite{macklin2021constraint}. 

Model reduction is another widely used acceleration technique, often referred to as the subspace method or reduced-order models. As the name implies, model reduction constructs a subspace representation of the fullspace DOFs. Modal analysis~\cite{pentland1989good, hauser2003interactive, choi2005modal} and its first-order derivatives~\cite{barbivc2005real} are commonly regarded as highly effective approaches for subspace construction. Additionally, displacements from recent fullspace simulations can be leveraged to enhance the subspace representation~\cite{kim2009skipping}. Some methods exploit condensation~\cite{teng2015subspace} and Schur complement~\cite{peiret2019schur}, which share the same nature of using a sub-set of DOFs to represent the global system status. \citet{sheth2015fully} addressed the momentum conservation among contacting reduced models. The success of deep learning also brings new perspectives to simulation. \citet{fulton2019latent} used an autoencoder to implicitly connect the latent space coordinate to the fullspace DOFs. \citet{shen2021high} employed complex-step finite difference (CSFD)~\cite{luo2019accelerated} to evaluate the fictitious force caused by varying subspaces. \citet{zong2023neural} built the subspace for the stress field instead of the displacement field.

A common drawback of reduced-order models lies in the lack of local details. As low-frequency deformations are normally considered more ``important'', and high-frequency deformations are therefore filtered by the subspace representation. This issue could be mitigated by building local subspaces. \citet{barbivc2011real} proposed a substructuring algorithm that assumes small and nearly rigid interfaces among local subspace domains, making it particularly effective for plant simulation~\cite{zhao2013interactive}. \citet{yang2013boundary} integrated modal warping~\cite{choi2005modal} with component mode synthesis (CMS)~\cite{macneal1971hybrid} to construct local subspaces based on interface deformations. To address locking artifacts, \citet{kim2011physics} introduced spring-based coupling between adjacent subspaces. Similarly, \citet{wu2015unified} utilized a spring-based coupling approach, combined with Cubature~\cite{an2008optimizing} sampling, to enhance efficiency and accuracy. \citet{harmon2013subspace} augmented the subspace with local bases to capture deformation induced by collision and contact.

Previous works have also utilized coarsened geometric representations to govern the dynamics of detailed models. For example, \citet{capell2002interactive} employed an embedded skeleton to deform elastic bodies, while \citet{gilles2011frame} used six-DOF rigid frames to drive deformable simulations. \citet{faure2011sparse} introduced scattered handles to model nonlinear dynamics, and \citet{lan2020medial,lan2021medial} leveraged the medial axis transform to construct mesh skeletons. \citet{martin2010unified} proposed sparsely distributed integrators, called elastons, to uniformly handle the nonlinear dynamics of rods, shells, and solids. These methods achieve significant speedups because the number of simulation DOFs is independent of the model's resolution. However, this comes at the cost of reduced accuracy and a loss of fine simulation details. After all, reduced simulation uses a low-dimension representation to model high-dimensional dynamics.

GPU simulation approaches the efficiency from a different perspective. Modern GPUs feature a large number of processors and excel in handling massive small-size computing tasks in parallel. This property requires an algorithmic re-design of simulation, shifting from a one-step solver (e.g., Newton's method) to parallelizable and iterative numerical procedures~\cite{fratarcangeli2018parallel}. For instance, \citet{wang2016descent} used Jacobi pre-conditioned gradient descent for elastic simulation. While the use of the Jacobi method increases the total number of iterations, the parallelized computation at the GPU compensates for it, resulting in improved overall performance. This idea can be combined with PD~\cite{wang2015chebyshev} to solve the global step system inexactly on the GPU. \citet{fratarcangeli2016vivace} used a parallel GPU GS method to solve the global step matrix, which shows a better convergence than Jacobi. GS has also been a popular choice for GPU-based XPBD implementations~\cite{macklin2016xpbd,chen2024position}.\citet{lan2022penetration} combined multiple Jacobi iterations into a single aggregated iteration named A-Jacobi. \citet{guo2024barrier} leveraged GPU for fast sparse matrix-vector computation to speed up nonlinear Newton Krylov solve. \citet{wu2022gpu} employed a multigrid-like pre-conditioner to further improve the convergence of GPU iterations. Similarly, subspace methods are also helpful to pre-condition the system~\cite{li2023subspace, lan2024efficient}. Despite the variety of simulation algorithm designs, GPU methods always trade convergence for parallelism, and achieving both superior convergence and parallelism is normally considered a ``mission impossible''. For example, to address the nonlinearity introduced by IPC (incremental potential contact) barriers~\cite{li2020incremental}, \citet{lan2023second} proposed a stencil descent method, which relaxes local variational energy at four vertices of an element (and a colliding primitive pair). In contrast, vertex block descent (VBD)~\cite{chen2024vertex} prioritizes parallelism by solving local problems at each vertex. As a result, stencil descent performs more effectively in simulations involving stiffer objects, while VBD is better suited for softer materials. 

%% file: overundershoot.tex
\newpage
\section{Undershoot \& Overshoot}\label{sec:overundershoot}
We start with an explanation of overshoot and show a second-order solution to this issue. Elastodynamic simulation can be formulated as a variational optimization at each discretized time step:
\begin{equation}\label{eq:variation}
    \arg \min_{\bm{x}} E(\bm{x}).
\end{equation}
The unknown vector $\bm{x} \in \mathbb{R}^N$ concatenates $x$, $y$, and $z$ coordinates of all the vertices of a finite element mesh, where $N$ denotes the system size. The target function $E = I + \Psi$ consists of the inertia potential $I$ and the elasticity potential $\Psi$, which penalize accelerated motions and mesh deformation, respectively. Suppose implicit Euler integration is used, $I$ becomes a quadratic function of $\bm{x}$:  $I =\frac{1}{2h^2}\|\bm{M}^{\frac{1}{2}}(\bm{x} - \bm{z})\|^2$, where $\bm{z} = \hat{\bm{x}} + h \hat{\dot{\bm{x}}} + h^2 \bm{M}^{-1} \bm{f}_{ext}$ is a known vector depending on the previous position $\hat{\bm{x}}$, velocity $\hat{\dot{\bm{x}}}$, and an external force $\bm{f}_{ext}$. $\bm{M}$ is the mass matrix, and $h$ is the time step size. $\Psi$ is a nonlinear function whose specific form depends on the chosen material model and the underlying constitutive law.  

We normally do not have a closed-form recipe to directly obtain $\bm{x}^\star$, the global minimizer of Eq.~\eqref{eq:variation}. Nearly all the nonlinear procedures start with an initial guess i.e., $\bm{x}^0$ and progressively improve this guess via $\bm{x}^{k+1} \leftarrow \bm{x}^k + \delta \bm{x}^k$. Here, the superscript $k$ denotes the iteration index. At each iteration, we would like to calculate an improving $\delta \bm{x}^k$ making $\bm{x}^{k+1}$ as close to $\bm{x}^\star$ as possible.

For GPU simulation, parallelization is commonly achieved by splitting $E$ into multiple sub-instances or sub-problems $E_i(\bm{x}_i)$. Here, the subscript $i$ is for the $i$-th sub-problem, which includes a small number of $N_i$ DOFs, $\bm{x}_i \in \mathbb{R}^{N_i}$. 
An extreme case is the coordinate descent (CD)~\cite{wright2015coordinate}, where each sub-problem contains just one unknown i.e., $N_i = 1$. When $N_i$ is small, sub-problems can be efficiently solved in parallel using Jacobi or GS scheme. Many GPU simulation algorithms widely used in the graphics community are designed following this high-level idea e.g., see~\cite{chen2024vertex,lan2023second}. 

A fundamental flaw of such a divide-and-conquer strategy is that \emph{minimizing $E_i$ does not align with minimizing $E$}. In other words, $\bm{x}^\star_i \neq \bm{S}_i\bm{x}^\star$, where $\bm{x}^\star_i$ is the local minimizer of $E_i$, and $\bm{S}_i$ is a selection matrix picking DOFs pertaining to the $i$-th sub-problem from the global vector. It may be possible that $\delta \bm{x}_i$ fails to sufficiently lower $E_i$, and therefore becomes less helpful reducing $E$. We refer to this issue \emph{undershoot}. Undershoot can be potentially alleviated with a \emph{local} line search with Wolfe condition~\cite{wolfe1969convergence}. Conversely, fully relaxing $E_i$ is also problematic because the reduction of $E_i$ often, if not always, fails to offset the energy increases when $\delta \bm{x}_i$ is applied. In other words, locally optimal $\delta \bm{x}_i$ can worsen the overall target function $E$, leading to the so-called \emph{overshoot}. Overshoot can only be monitored with global line search, which is expensive and should not be frequently used. Overshoot suggests the local computation is inaccurate since it only uses local information.

%% file: second.tex
\section{Towards Second-order Convergence}\label{sec:second}
The ideal situation free of overshoot (and undershoot) occurs when the local solve update yields $\delta \bm{x}^{k+1}_i = \bm{S}_i \delta \bm{x}^\star$. This means that the system converges with one single iteration. Yet it is unlikely because $\delta \bm{x}^\star$ is unknown, and it is next to impossible to steer the local solve towards an uncharted target. We take a step back and aim to push the local solve to achieve global second-order convergence, i.e., at a similar rate to Newton's method.

Newton's method is well-known, which Taylor expands $E(\bm{x}^\star)$ at $\bm{x}^k$ such that:
\begin{multline}\label{eq:newton}
    E(\bm{x}^\star) = E^\star = E(\bm{x}^k + \delta \bm{x}^k) \\
     = E^k + \bm{g}^{k^\top} \delta \bm{x}^k + \frac{1}{2} \delta \bm{x}^{k^\top} \bm{H}^k \delta \bm{x}^k + O(\|\delta \bm{x}^k\|^3),
\end{multline}
where $E^k = E(\bm{x}^k)$. $\bm{g}^k = \left(\frac{\partial E^k }{ \partial \bm{x}}\right)^\top \in \mathbb{R}^N$, and $\bm{H}^k = \frac{\partial^2 E^k}{\partial \bm{x}^2} \in \mathbb{R}^{N \times N}$ are the gradient and Hessian of the variational energy $E$. If $O(\|\delta \bm{x}^k\|^3)$ is sufficiently small, and $E$ is secondary differentiable, Newton's method converges quadratically without needing the line search~\cite{nocedal1999numerical}. The corresponding DOF update $\delta \bm{x}^*$ can be computed from:
\begin{equation*}\label{eq:newton_approximate}
 \delta \bm{x}^* = \arg \min_{\bm{y}} E(\bm{y}) =  E^k + \bm{g}^{k^\top} \bm{y} + \frac{1}{2} \bm{y}^\top \bm{H}^k \bm{y} 
\end{equation*}
via solving the linear system of:
\begin{equation}\label{eq:newton_system}
    \bm{H}^k \delta \bm{x}^* = -\bm{g}^k.
\end{equation}
Eq.~\eqref{eq:newton_system} gives a closed-form way to compute $\delta \bm{x}^*$ so that the update $\bm{x}^* \leftarrow \bm{x}^k + \delta \bm{x}^*$ offers the second-order optimal estimation of $\bm{x}^\star$. If we make the local solve $\delta\bm{x}_i$ approach $\bm{S}_i \delta\bm{x}^*$, a parallel iteration becomes as converging as a (global) Newton step.

Let $E_{C_i}$ be the complement of $E_i$ i.e., the variational energy at the remaining parts of the model such that $E_{C_i} = E -  E_i$. Since overshoot stems from the lack of global information, a straightforward remedy is to change the local sub-problem to:
\begin{equation}\label{eq:ideal_local}
    \min_{\delta x_i} E_i(\delta \bm{x}_i) + E_{C_i}(\delta \bm{x}),
\end{equation}
so that the local solve also takes the global energy variation into account. It should be immediately noticed that $E_{C_i}(\delta \bm{x})$ depends on $\delta \bm{x}$ while the variable to be optimized in Eq.~\eqref{eq:ideal_local} only involves local DOFs $\delta \bm{x}_i$. To make Eq.~\eqref{eq:ideal_local} meaningful, we can choose to change the unknown from $\delta \bm{x}_i$ to $\delta \bm{x}$, which essentially converts Eq.~\eqref{eq:ideal_local} back to Eq.~\eqref{eq:variation} --- we give up the parallelization for the convergence. 

Alternatively, if we know how $\delta \bm{x}_i$ would influence the global DOF variation $\delta \bm{x}$ such that $\delta \bm{x} = \phi_i(\delta \bm{x}_i)$, Eq.~\eqref{eq:ideal_local} becomes:
\begin{equation}\label{eq:app_local}
    \min_{\delta x_i} E_i(\delta \bm{x}_i) + E_{C_i} \left[ \phi_i(\delta \bm{x}_i) \right],
\end{equation}
which can then be solved e.g., using Newton's method as:
\begin{equation}\label{eq:local_newton}
    \delta \bm{x}_i = -\left(\bm{H}_i +  \nabla E_{C_i}^\top \frac{\partial^2 \phi_i}{\partial \delta \bm{x}^2_i} + \frac{\partial \nabla E_{C_i}}{\partial \delta \bm{x}_i}^\top \frac{\partial \phi_i}{\partial \delta \bm{x}_i} \right)^{-1}  \left(\bm{g}_i +  \frac{\partial \phi_i}{\partial \delta \bm{x}_i}^\top \nabla E_{C_i} \right),
\end{equation}
where $\nabla E_{C_i} = \left(\frac{\partial E_{C_i}}{\partial \delta\bm{x}}\right)^\top = \left(\frac{\partial E_{C_i}}{\partial \bm{x}} \right)^\top \in \mathbb{R}^N$, and $\frac{\partial \nabla E_{C_i}}{\partial \delta \bm{x}_i} = \frac{\partial \nabla E_{C_i}}{\partial \bm{x}_i} = \frac{\partial^2 E_{C_i}}{\partial \bm{x}\partial \bm{x}_i} \in \mathbb{R}^{N \times N_i}$. Intuitively, $\phi_i(\delta \bm{x}_i)$ describes the global deformation increment caused by the local perturbation of $\delta \bm{x}_i$.

\subsection{Local Perturbation Subspace}
At the current Newton linearization, we compute $\phi_i$ by imposing a unit perturbation at one local DOF while keeping all the other local DOFs fixed. The resulting perturbation at the rest part of the model represents the influence of the perturbed local DOF. To this end, we re-order system DOFs as $\delta \bm{x} = [\delta \bm{x}_i^\top, \delta \bm{x}_{C_i}^\top]^\top$ such that $\delta \bm{x}_{C_i} \in \mathbb{R}^{N - N_i}$ contains all the \emph{complementary} DOFs excluding the ones in $\delta \bm{x}_i$. We can then build $N_i$ \emph{incremental equilibria}:
\begin{equation}\label{eq:cms}
    \left[
    \begin{array}{cc}
    \bm{H}_{i,i} & \bm{H}_{i, C_i} \\
    \bm{H}^\top_{i, C_i} & \bm{H}_{C_i, C_i}
    \end{array}
    \right]
    \left[
    \begin{array}{c}
    \bm{I}\\
    \bm{U}_{C_i}
    \end{array}
    \right]
    =
    \left[
    \begin{array}{c}
    \delta \bm{F}_i\\
    \bm{0}
    \end{array}
    \right].
\end{equation}
Note that the superscript $k$ is ignored in Eq.~\eqref{eq:cms}. $\bm{I}$ is an $N_i$ by $N_i$ identity matrix. $\delta \bm{F}_i$ are the virtual forces needed to trigger the unit perturbation at each local DOF and keep others fixed. Columns in $\bm{U}_{C_i}$ embody the corresponding virtual deformation at complementary DOFs, which can be computed by expanding the second row of Eq.~\eqref{eq:cms}:
\begin{equation}
    \bm{H}^\top_{i, C_i} + \bm{H}_{C_i, C_i} \bm{U}_{C_i} = \bm{0} \Rightarrow \bm{U}_{C_i} = -\bm{H}^{-1}_{C_i, C_i} \bm{H}^\top_{i, C_i}.
\end{equation}

Any local solve update $\delta \bm{x}_i$ can be understood as a linear combination of such per-DOF perturbations i.e., $\delta \bm{x}_i = \bm{I} \delta \bm{x}_i$. Consequently, the triggered global perturbation is also a linear combination of columns in $\bm{U}_{C_i}$. In other words, $\bm{U}_{C_i}$ forms a set of bases spanning a perturbation subspace at complementary DOFs, in which the perturbations are controlled by $\delta \bm{x}_i$. Therefore, $\phi$ can be obtained as:
\begin{equation}\label{eq:phi}
    \delta \bm{x} = 
    \phi_i(\delta \bm{x}_i)= 
    \left[
    \begin{array}{c}
    \delta \bm{x}_i\\
    \delta \bm{x}_{C_i}
    \end{array}
    \right]
    =
    \left[
    \begin{array}{c}
    \bm{I}\\
    -\bm{H}^{-1}_{C_i, C_i} \bm{H}^\top_{i, C_i}
    \end{array}
    \right]
    \delta \bm{x}_i
    =
    \left[
    \begin{array}{c}
    \bm{I}\\
    \bm{U}_{C_i}
    \end{array}
    \right]
    \delta \bm{x}_i.
\end{equation}
We note that $\phi_i$ becomes linearized at the current Newton step. It remains a nonlinear function during the simulation because the Hessian $\bm{H}(\bm{x}^k)$ depends on the current deformation of the system. 

\subsection{Optimality of $\phi_i$}
We argue that Eq.~\eqref{eq:phi} builds the optimal local subspace so that $\delta \bm{x}_i$ computed using Eq.~\eqref{eq:local_newton} matches the global Newton solve $\bm{S}_i \delta \bm{x}^*$. To see this, we re-organize Eq.~\eqref{eq:newton_system} in a similar way:
\begin{equation}\label{eq:blocl_newton}
    \left[
    \begin{array}{cc}
    \bm{H}_{i, i} & \bm{H}_{i, C_i} \\
    \bm{H}^\top_{i, C_i} & \bm{H}_{C_i, C_i}
    \end{array}
    \right]
    \left[
    \begin{array}{c}
    \delta \bm{x}^*_i\\
    \delta \bm{x}^*_{C_i}
    \end{array}
    \right]
    =
    \left[
    \begin{array}{c}
    -\bm{g}_i\\
    -\bm{g}_{C_i}
    \end{array}
    \right].
\end{equation}
Expanding the second line, we can have:
\begin{equation}\label{eq:xr}
    \delta \bm{x}^*_{C_i} = -\bm{H}^{-1}_{C_i, C_i}\left( \bm{g}_{C_i} + \bm{H}^\top_{i, C_i} \delta \bm{x}^*_i \right).
\end{equation}
Substituting Eq.~\eqref{eq:xr} back to the first line of Eq.~\eqref{eq:blocl_newton} yields:
\begin{align}\label{eq:deltaxi}
   & \quad \bm{H}_{i,i}\delta \bm{x}^*_i + \bm{H}_{i, C_i} \delta \bm{x}^*_{C_i} = -\bm{g}_i \nonumber\\
&\Rightarrow  \bm{H}_{i,i} \delta \bm{x}^*_i - \bm{H}_{i, C_i} \bm{H}^{-1}_{C_i, C_i}\left( \bm{g}_{C_i} + \bm{H}^\top_{i, C_i} \delta \bm{x}^*_i \right) = -\bm{g}_i \nonumber\\
&\Rightarrow   \left(\bm{H}_{i,i} - \bm{H}_{i, C_i} \bm{H}^{-1}_{C_i, C_i}\bm{H}^\top_{i, C_i}\right) \delta \bm{x}^*_i = \bm{H}_{i, C_i} \bm{H}^{-1}_{C_i, C_i}\bm{g}_{C_i} -\bm{g}_i \nonumber\\
&\Rightarrow  \delta \bm{x}^*_i = \left( \bm{H}_{i,i} + \bm{H}_{i,C_i} \bm{U}_{C_i} \right)^{-1} \left(\bm{H}_{i, C_i} \bm{H}^{-1}_{C_i, C_i}\bm{g}_{C_i} -\bm{g}_i \right).
\end{align}

Meanwhile, we also have the following relations:
\begin{equation}\label{eq:relation}
    \left\{
    \begin{array}{ll}
\displaystyle \frac{\partial \phi_i}{\partial \delta \bm{x}_i} = 
    \left[
    \begin{array}{c}
    \bm{I}\\
    \bm{U}_{C_i}
    \end{array}
    \right], &
    \displaystyle
    \frac{\partial^2 \phi_i}{\partial \delta \bm{x}^2_i} = \bm{0}, \\
    \displaystyle
    \nabla E_{C_i} = 
    \left[
    \begin{array}{c}
    \bm{0} \\
    \bm{g}_{C_i}
    \end{array}
    \right], &
    \displaystyle
    \frac{\partial \nabla E_{C_i}}{\partial \delta \bm{x}_i} = \frac{\partial^2 E_{C_i}}{\partial \bm{x}\partial \bm{x}_i} =
    \left[
    \begin{array}{c}
    \bm{0} \\
    \bm{H}^\top_{i, C_i}
    \end{array}
    \right].
    \end{array}
    \right.
\end{equation}
Note that $\bm{H}_i$ in Eq.~\eqref{eq:local_newton} is just $\bm{H}_{i, i}$ in Eqs.~\eqref{eq:cms} and \eqref{eq:blocl_newton}. Putting them together with  Eq.~\eqref{eq:relation} back to Eq.~\eqref{eq:local_newton}, we obtain:
\begin{equation*}
 \delta \bm{x}_i = \left( \bm{H}_{i, i} + \bm{H}_{i, C_i} \bm{U}_{C_i} \right)^{-1} \left(\bm{H}_{i, C_i} \bm{H}^{-1}_{C_i, C_i}\bm{g}_{C_i} -\bm{g}_i \right).
\end{equation*}
This shows that the local optimization using Eqs.~\eqref{eq:local_newton} and \eqref{eq:phi} is mathematically equivalent to solving the global Newton and satisfies $\delta \bm{x}_i = \bm{S}_i \bm{x}^*$. In other words, the resulting $\delta \bm{x}_i$ is second-order optimal.

\subsection{Co-rotated Subspace}
Computing $\phi^k_i$ via Eq.~\eqref{eq:phi} needs the current Hessian matrix $\bm{H}^k(\bm{x}^k)$, which varies under different mesh poses. Clearly, it is infeasible to re-build $\phi_i^k$ at each time step.

Recall that $\phi_i$ helps avoid overshoot because it brings awareness of $E_{C_i}$ during the local solve. That said, the key to mitigating overshoot lies in how well $E_{C_i}(\phi_i^k)$ is estimated. This requirement is less stringent than demanding $\phi_i^k = \delta \bm{x}^*$. The latter always guarantees that $E_{C_i}(\phi^k_i)$ is exact; the reverse, however, does not always hold true. Therefore, it suffices to construct alternative subspace function $\tilde{\phi}_i^k$ such that $E_{C_i}(\tilde{\phi}_i^k)$ closely matches $E_{C_i}(\phi_i^k)$ even though $\tilde{\phi}_i^k(\delta \bm{x}_i)$ may not exactly align with $\phi_i^k(\delta \bm{x}_i)$.

The idea is to embed a co-rotated local frame at each mesh vertex. Given the current deformation pose $\bm{x}^k$, we extract a local rotation $\bm{R}^k_{\langle j \rangle} \in {SO}(3)$ at vertex $j$. Here, we use the subscripted notation $\langle j \rangle$ to denote the vertex index. $\bm{R}^k_{\langle j \rangle}$ can be efficiently computed by applying the polar decomposition over the local deformation gradient at all the vertices in parallel. This local rotation captures how an infinitesimal chunk of material around the vertex $j$ is rotated w.r.t. its rest pose. As a rigid rotation preserves the elasticity energy, we rotate the vertex back to this rest orientation without modifying the energy it stores. Meanwhile, $\phi$ at the rest shape can be pre-computed. Mathematically, this strategy builds $\tilde{\phi}_i^k$ as:
\begin{equation}\label{eq:tilde_phi}
\tilde{\phi}^k_i = 
\bm{R}^k
\underbrace{
\left[
\begin{array}{c}
\bm{I}\\
-\bar{\bm{H}}^{-1}_{C_i, C_i} \bar{\bm{H}}^\top_{i, C_i}
\end{array}
\right]
}_{\bar{\bm{U}}_i}
\bm{R}_i^{k^\top}
\delta \bm{x}^k_i
=
\underbrace{
\bm{R}^k
\bar{\bm{U}}_i \bm{R}_i^{k^\top} 
}_{{\bm{U}}^k_i}
\delta \bm{x}^k_i,
\end{equation}
where $\bm{R}^k $ and $\bm{R}^k_i$ are two block-diagonal matrices whose 3 by 3 diagonal blocks are per-vertex local rotation matrix. At the current Newton step, $\bm{R}^k $ and $\bm{R}^k_i$ are constant. $\bar{\bm{H}}^{-1}_{C_i, C_i} \bar{\bm{H}}^\top_{i, C_i}$ is also constant depending on the rest-shape Hessian $\bar{\bm{H}}$. In other words, $\tilde{\phi}_i^k$ remains a linearized subspace at the current Newton linearization, and it can also be efficiently constructed at different deformation poses because $\bar{\bm{H}}^{-1}_{C_i, C_i}\bar{\bm{H}}^\top_{i, C_i}$ is pre-computable.

\subsection{Discussion}
The subspace function $\phi_i^k$ plays a central role. It allows global awareness by predicting how the global energy changes in response to the local update $\delta \bm{x}_i$. As a result, the local solve well aligns with $\bm{S}_i\delta \bm{x}^*$. $\phi_i^k$ does not reduce the size of the local problem so that the local solve remains in the $N_i$-rank space. However, the remainder part of the optimization ($E_{C_i}$) is condensed to a subspace whose generalized subspace coordinate is designed to be $\delta \bm{x}_i$. While this is a nonlinear subspace that varies w.r.t. deformation poses, it can be linearized at each local quadratic approximation of $\bm{x}^\star$. We design an alternative formulation $\tilde{\phi}_i^k$ to allow pre-computation for the most expensive part of the subspace construction. As a result, the reduced Hessian of $\frac{\partial^2 E_{C_i}}{\partial \delta \bm{x}_i^2}$ acts as a ``damper'' preventing the local solver from reaching its local minimizer (and thus overshoots).


The subspace function $\phi_i^k$ can also be viewed as a type of interpolation function that smoothly propagates $\delta \bm{x}_i$ to $\delta \bm{x}$. To this end, many interpolation algorithms may also help such as radial basis function (RBF)~\cite{botsch2005real,carr2001reconstruction}, Green coordinates~\cite{lipman2008green,michel2023polynomial}, Splines~\cite{li2023subspace,liu2014skinning}, Harmonics~\cite{lipman2010biharmonic,jacobson2011bounded}, or SPH kernels~\cite{koschier2022survey}. However, those methods are geometry-based and do not reflect the material property. For instance, $\delta \bm{x}_i$ produces more general and global perturbations for stiff materials and more local and regional perturbations for soft materials. The perturbations at $x$, $y$, or $z$ coordinates are also different given different constitutive models. After all, geometry-based interpolation schemes are not designed to make local solve second-order optimal i.e., $\delta \bm{x}_i = \bm{S}_i \delta \bm{x}^*$. On the other hand, $\phi$ performs like a material-aware shape function, expanding the influence of a sub-problem towards the entire object.

%% file: cubature.tex
\section{Cubature Sampling}\label{sec:cubature}
$\tilde{\phi}_i^k$ is linearized at the current deformation poses $\bm{x}^k$, and we have $\frac{\partial \tilde{\phi}^k_i}{\partial \delta \bm{x}_i} = \bm{U}^k_i$ and $\frac{\partial^2 \tilde{\phi}^k_i}{\partial \delta \bm{x}^2_i} = \bm{0}$ per Eq.~\eqref{eq:tilde_phi}. Substituting them into Eq.~\eqref{eq:local_newton} with some manipulations gives the local system we need to solve:
\begin{equation}\label{eq:reduced_local}
    \left( \bm{H}^k_{i, i} + \tilde{\bm{H}}^k_{i, i} \right) \delta \bm{x}^k_i = -\tilde{\bm{g}}^k_i - \bm{g}^k_i,
\end{equation}
where 
\begin{equation}\label{eq:reduced_h_g}
    \tilde{\bm{H}}^k_{i, i} = \bm{U}^{k^\top}_i \left(\nabla^2 E^k_{C_i}\right) \bm{U}^k_i,
    \quad 
    \tilde{\bm{g}}^k_i = \bm{U}^{k^\top}_i \nabla E^k_{C_i},
\end{equation}
are the reduced Hessian and gradient force. Eq.~\eqref{eq:reduced_local} is of low dimension and can be efficiently solved at each GPU thread. However, its assembly is not, since $\bm{U}_i^k$ is a dense matrix. To exactly build $\tilde{\bm{H}}^k_{i, i}$ and $\tilde{\bm{g}}^k_i$, one needs to traverse all the complementary DOFs and project their Hessian and gradient into the column space of $\bm{U}^k_i$. The time complexity is $O(N \cdot N_i^2)$, and it needs to be done for all the sub-problems.

A known solution was proposed in \cite{an2008optimizing} a.k.a. Cubature. Cubature is a sampling technique, which pre-computes a small group of sample elements $\mathcal{S}_i$ i.e., Cubature elements, and the associated non-negative weights. The reduced Hessian and gradient force are then approximated by:
\begin{equation}\label{eq:cubature}
    \tilde{\bm{H}}^k_{i, i} \approx \sum_{e\in\mathcal{S}_i} w_e [\bm{U}^k_i]^\top_e [\nabla^2 E^k_{C_i} ]_e [\bm{U}^k_i]_e,\; \tilde{\bm{g}}^k_i \approx \sum_{e\in\mathcal{S}_i} w_e [\bm{U}^k_i]^\top_e [\nabla E^k_{C_i}]_e,
\end{equation}
where $[\nabla E^k_{C_i}]_e \in \mathbb{R}^{12}$ and $[\nabla^2 E^k_{C_i} ]_e \in \mathbb{R}^{12 \times 12}$ are the gradient and Hessian of $E^k_{C_i}$ at the Cubature element $e$. $[\bm{U}^k_i]_e \in \mathbb{R}^{12 \times N_i}$ is the element subspace matrix, which extracts corresponding rows of $e$ from $\bm{U}^k_i$. $w_e$ is the non-negative weight coefficient.

Given a set of training poses $\mathcal{T}$, the corresponding reduced gradient (i.e., $\tilde{\bm{g}}_i^{(1)}$,..., $\tilde{\bm{g}}_i^{(|\mathcal{T}|)}$), and Cubature element set $\mathcal{S}_i$, the weight coefficients at all the Cubature elements are computed via solving:
\begin{equation}\label{eq:cubature_weight}
    \left[
    \begin{array}{ccc}
    \frac{[\tilde{\bm{g}}_i^{(1)}]_1}{\|\tilde{\bm{g}}_i^{(1)}\|} & \cdots & \frac{[\tilde{\bm{g}}_i^{(1)}]_{|\mathcal{S}_i|}}{\|\tilde{\bm{g}}_i^{(1)}\|}\\
    \vdots & \cdots & \vdots \\
    \frac{[\tilde{\bm{g}}_i^{(|\mathcal{T}|)}]_1}{\|\tilde{\bm{g}}_i^{(|\mathcal{T}|)}\|} & \cdots & \frac{[\tilde{\bm{g}}_i^{(|\mathcal{T}|)}]_{|\mathcal{S}_i|}}{\|\tilde{\bm{g}}_i^{(|\mathcal{T}|)}\|}
    \end{array}
    \right]
    \left[
    \begin{array}{c}
    w_1 \\
    \vdots\\
    w_{|\mathcal{S}|_i}
    \end{array}
    \right]
    =
    \left[
    \begin{array}{c}
    \frac{\tilde{\bm{g}}_i^{(1)}}{\|\tilde{\bm{g}}_i^{(1)}\|} \\
    \vdots \\
    \frac{\tilde{\bm{g}}_i^{(|\mathcal{T}|)}}{\|\tilde{\bm{g}}_i^{(|\mathcal{T}|)}\|}
    \end{array}
    \right],
\end{equation}
using non-negative least square (NNLS). Several candidate Cubature elements are randomly picked from the non-Cubature elements, and the one most effectively reduces the residual of Eq.~\eqref{eq:cubature_weight} will be included in $\mathcal{S}_i$. This procedure continues until the desired residual error is reached. We refer the reader to the seminal paper of Cubature~\cite{an2008optimizing} for further details.

Cubature training is normally considered expensive, and acceleration techniques are also available e.g., see~\cite{yang2015expediting,von2013efficient}. The complexity of Cubature training is largely due to repetitive NNLS solving, which scales up super-polynomially when more Cubature samples are needed. The size of Cubature sample set is linearly correlated with the size of the subspace i.e., $|\mathcal{S}_i| \varpropto N_i$. As a result, we only need a handful Cubature elements e.g., four or six, for each sub-problem in our implementation (with residual less than $1\%$), making Cubature training lightweight.  

We would also like to mention that unlike most existing algorithms for reduced simulation, which aim to construct a compact subspace for \emph{deformation} or \emph{displacement}, our subspace depicts deformable \emph{perturbation} $\delta \bm{x}$. Therefore, $\bm{U}^k_i$ does not need to incorporate large deformation. To this end, our training poses are simply low-frequency eigenvectors of the rest-shape Hessian matrix $\bar{\bm{H}}$. In addition, what we need is a good estimation of $E_{C_i}$. The exactness of Cubature gradient is of less importance -- this is because our primary goal is to estimate a reasonable ``damping Hessian'' to prevent overshoot other than exactly minimize $E$. As a result, sparse Cubature sampling is highly effective in our framework.




%% file: pre_computation.tex
\section{Full-coordinate Pre-computation}\label{sec:pre_computation}
The co-rotated formulation allows the most expensive step in constructing  $\tilde{\phi}^k_i$ to be pre-computed as shown in Eq.~\eqref{eq:tilde_phi}. For each sub-problem $i$, the pre-computation solves $\bar{\bm{H}}_{C_i, C_i}
$ for $N_i$ times. $\bar{\bm{H}}_{C_i, C_i}$ is a $(N - N_i)\times(N - N_i)$ matrix. It is nearly of the same scale as $\bar{\bm{H}}$ since $N_i$ is a small quantity. Performing such factorization for all the sub-problems is extremely slow, which takes days for large-scale models. We observe that while $\bar{\bm{H}}_{C_i, C_i}$ differs across different sub-problems, a significant portion of these matrices overlaps. This suggests a brute-force computation would be inefficient and wasteful. Motivated by this observation, we propose a full-coordinate formulation for Eq.~\eqref{eq:cms}, which accelerates the pre-computation by \emph{three orders}, reducing the time required from days to tens of minutes.

We know that Eq.~\eqref{eq:cms} builds $\tilde{\phi}^k_i$ via a set of incremental equilibria. Each column of $\delta \bm{F}_i$ embodies an external force at local DOFs, which triggers a unit perturbation at a specific local DOF while keeping other local DOFs fixed. The very same equilibrium can also be achieved by directly prescribing local DOF values with $N_i$ position constraints:
\begin{equation}\label{eq:cms_constraint}
    \bar{\bm{H}} \bar{\bm{u}}_{i, j} = \bm{0},\quad \text{s.t.}\; \bm{S}_i \bar{\bm{u}}_{i, j} = \bm{e}_j,\, \text{for}\, j = 1,2,\cdots,N_i,
\end{equation}
where $\bm{e}_j$ is the $j$-th column of $\bm{I}$, which is a vector of zeros with a value of one at the $j$-th entry. $\bar{\bm{u}}_{i, j} \neq \bm{0}$ is the deformation variation of the mesh induced by the constraint. The subscript $i, j$ suggests constraints are imposed for the $i$-th sub-problem, whose $j$-th local DOF is perturbed. This constrained linear system can be solved with the full coordinate and Lagrange multipliers such that:
\begin{equation}\label{eq:cms_multiplier}
\left[
\begin{array}{cc}
\bar{\bm{H}} & \bm{S}_i^\top \\
\bm{S}_i & \bm{0}
\end{array}
\right]
\left[
\begin{array}{c}
\bar{\bm{U}}_i \\
\bm{\Lambda}_i 
\end{array}
\right]
=
\left[
\begin{array}{c}
\bm{0} \\
\bm{I} 
\end{array}
\right].
\end{equation}
Here, $\bar{\bm{U}}_i = [\bar{\bm{u}}_{i, 1},\cdots,\bar{\bm{u}}_{i, N_i}]$. $\bm{\Lambda}_i  = -\delta \bm{F}_i$ are Lagrange multipliers. Compared with Eq.~\eqref{eq:cms}, Eq.~\eqref{eq:cms_multiplier} solves both unknown DOFs and multipliers, and it is normally considered less efficient. However, because its top-left block becomes invariant for all the sub-problems, we can leverage block-wise matrix inversion to re-use factorized $\bar{\bm{H}}$ without performing factorization of different $\bar{\bm{H}}_{C_i, C_i}$ repeatedly.

The block-inverse of the l.h.s. of Eq.~\eqref{eq:cms_multiplier} gives:
\begin{multline}\label{eq:block_inv}
 \left[
\begin{array}{cc}
\bar{\bm{H}} & \bm{S}_i^\top \\
\bm{S}_i & \bm{0}
\end{array}
\right]^{-1} 
= 
\left[
\begin{array}{cc}
\bar{\bm{H}}^{-1} + \bar{\bm{H}}^{-1} \bm{S}_i^\top \bar{\bm{G}}_i \bm{S}_i \bar{\bm{H}}^{-1}  
& -\bar{\bm{H}}^{-1} \bm{S}_i^\top \bar{\bm{G}}_i \\
- \bar{\bm{G}}_i \bm{S}_i \bar{\bm{H}}^{-1} 
& \bar{\bm{G}}_i
\end{array}
\right],
\end{multline}
where $\bar{\bm{G}}_i = -\left(\bm{S}_i \bar{\bm{H}}^{-1} \bm{S}_i^\top\right)^{-1}$ is the inverse of Schur complement of $\bar{\bm{H}}$. Inverting Eq.~\eqref{eq:cms_multiplier} leads:
\begin{equation}
    \left[
\begin{array}{c}
\bar{\bm{U}}_i \\
\bm{\Lambda}_i 
\end{array}
\right] =  \left[
\begin{array}{cc}
\bar{\bm{H}} & \bm{S}_i^\top \\
\bm{S}_i & \bm{0}
\end{array}
\right]^{-1} 
\left[
\begin{array}{c}
\bm{0} \\
\bm{I} 
\end{array}
\right],
\end{equation}
which shows that:
\begin{equation}\label{eq:u}
    \bar{\bm{U}}_i = -\bar{\bm{H}}^{-1} \bm{S}_i^\top \bar{\bm{G}}_i.
\end{equation}
During the pre-computation, we only factorize $\bar{\bm{H}}$ once, which is shared by all the sub-problems without factorizing different $\bar{\bm{H}}_{C_i, C_i}$ at individual sub-problems. $\bar{\bm{G}}_i \in \mathbb{R}^{N_i \times N_i}$ is a small matrix and can be efficiently calculated by solving $\bar{\bm{H}}$ for $N_i$ times. As a result, the pre-computation of a sub-problem only needs $2N_i$ forward and backward substitutions of $\bar{\bm{H}}$, and such computation can be trivially parallelized at all the sub-problems.

%% file: collision.tex
\section{Incremental Potential Contact }
Collisions among deformable objects can also be considered as a type of potential energy and uniformly encoded in the variational optimization. A representative paradigm is the incremental potential contact or IPC~\cite{li2020incremental}. IPC is a primal implementation of the interior-point method, which injects a logarithmic barrier energy into Eq.~\eqref{eq:variation}, defined at each surface primitive pair $l$:
\begin{equation}\label{eq:barrier}
    B_l(d_l, \hat{d}) =
    \left\{
    \begin{array}{ll}
    -(d_l - \hat{d})^2 \log \left(\frac{d_l}{\hat{d}}\right), & 0 < d_l < \hat{d}\\
    0, & d_l \geq \hat{d}
    \end{array}
    \right..
\end{equation}
$\hat{d}$ is a user-specified parameter prescribing the collision tolerance. $B_l$ becomes ``active'' if the closest distance between a collision pair (i.e., $d_l$) is smaller than $\hat{d}$, and it approaches to $+\infty$ as $d_l$ approaches to $0$. Intuitively, IPC offers a nonlinear penalty mechanism pushing a pair of colliding primitives, e.g., a vertex-triangle pair or an edge-edge pair, away from each other when they are in proximity. 


Our method approaches improved convergence from an optimization point of view. As a result, it is compatible with IPC or other implicit penalty methods as long as the collision resolution is in the form of an unconstrained optimization. The key adaptation is to accommodate $\phi_i$ with collisions and contacts. When a vertex $V$ on model $A$ is in contact with another object $B$ $\phi_i$ does not only concern the total energy on $A$ but also has a non-vanishing influence on the energy on $B$. In other words, $A$ and $B$ become two-way coupled by the contact at the vertex. The value of the subspace basis on $A$ is pre-computed. While the values of $\phi_i$  at $B$ can be approximated by assuming all the colliding vertices on $B$ have the same perturbation as $V$. This strategy assumes IPC or the contact penalty is much stiffer than the elasticity stiffness, and the variation of the perturbation within the sub-problem is ignored.

%% file: exp.tex
\section{Experimental Results}\label{sec:exp}
We implemented our pipeline on a desktop computer with an \textsf{intel} \textsf{i7-12700} CPU (for pre-computation) and an \textsf{Nvidia} \textsf{3090 RTX} GPU. We used \textsf{Spectra} library for computing the eigendecomposition of the rest-shape Hessian matrix $\bar{\bm{H}}$. It should be noted that our framework can be conveniently deployed with other parallel computing platforms, such as multi-core CPUs. Nevertheless, this section reports the simulation performance and results based on our GPU implementation. Tab.~\ref{tab:time} lists detailed experiment setups and timing information. For all the examples, we normalize the scene into a one-by-one-by-one unit cube and use the change of position of the object between two consecutive iterations i.e., $\| \Delta \bm{x} \|$ as a simple and uniformed measure for convergence check. Please also refer to the supplementary video for more animation results.  

\begin{table*}
\caption{\textbf{Experiment statistics.}~~This table reports time statistics and simulation setups for all the experiments mentioned in the paper. \textbf{$\#$ Element} gives the total number of elements in the example. \textbf{$\#$ DOF} is the total number of simulation DOFs. \textbf{$|\mathcal{S}_i|$} is the size of Cubature samples used for each sub-problem, knowing that each sub-problem has three DOFs. \textbf{$\|\Delta \bm{x}\|$} is the convergence condition. \textbf{$h$} gives the time step size used for the simulation. \textbf{Parallelism} tells if the parallelization is in a Jacobi way or a GS way i.e., Jacobi or GS. The difference between those two parallelization methods is negligible. The column of \textbf{Collision} shows what method is used for collision resolution, using either the implicit penalty method (\textcolor{blue}{Penalty}) or incremental potential contact (\textcolor{orange}{IPC}). \textbf{$\#$ Iteration} reports the average number of iterations needed to simulate a time step. \textbf{Pre. time} is the total time used for pre-computation, and \textbf{Sim. time} is the average simulation time for simulating one time step. We also report the acceleration rate our method offers compared with VBD~\cite{chen2024vertex} if the collision is processed with the penalty method or GPU-IPC~\cite{guo2024barrier} if the collision is processed with the IPC barrier. We use stable Neo-Hookean model~\cite{smith2018stable} for all the deformable body experiments. For cloth simulation result i.e., Fig.~\ref{fig:cloth}, we use StVK model to capture the in-plane deformation, and quadratic bending for the out-of-plane deformation.}\label{tab:time}
{\small 
\begin{center}
\def\arraystretch{1.1}
\begin{tabular}{|c||c|c|c|c|c|c|c|c|c|c|}
\whline{1.15pt}
\textbf{Scene} & \textbf{$\#$ Element} & \textbf{$\#$ DOF} & \textbf{$|\mathcal{S}_i|$}& \textbf{$h$} & \textbf{$\|\Delta \bm{x}\|$} & \textbf{Parallelism} & \textbf{Collision} & \textbf{$\#$ Iteration} & \textbf{Pre. time} & \textbf{Sim. time} \\
\whline{0.5pt}

Teaser (Fig.~\ref{fig:teaser}) & $3.5$M & $4.4$M & $4$ & $1/120$ & $1E-3$ & GS & \textcolor{blue}{Penalty} & $55$ & $37$ min. & $855$ ms~(\textcolor{blue}{$\infty\times$})\\

Falling Armadillos (Fig.~\ref{fig:dropingAarmadillo}) & $6$M & $3.6$M & $4$ & $1/150$ & $5E-4$ & Jacobi & \textcolor{blue}{Penalty} & $58$ & $52$ min. & $883$ ms~(\textcolor{orange}{$78\times$})\\

House of cards (Fig.~\ref{fig:cardhouse}) & $394$K & $372$K & $4$ & $1/50$ & $1E-3$ & Jacobi & \textcolor{orange}{IPC} & $23$ & $1$ min. & $31$ ms~(\textcolor{orange}{$120\times$})\\

Dragon (Fig.~\ref{fig:realtime}) & $100$K & $80$K & $6$ & $1/100$ & $1E-3$ & Jacobi & \textcolor{blue}{Penalty} & $9$ & $7$ min. & $7.3$ ms~(\textcolor{blue}{$32\times$})\\

Letters soft (Fig.~\ref{fig:letter}) & $2.1$M & $1.7$M & $6$ & $1/120$ & $5E-4$ & GS & \textcolor{blue}{Penalty} & $27$ & $37$ min. & $176$ ms~(\textcolor{blue}{$43\times$})\\

Barbarian ships (Fig.~\ref{fig:ship}) & $2.5$M & $2.1$M & $4$ & $1/120$ & $3E-4$ & Jacobi & \textcolor{blue}{Penalty} & $34$ & $45$ min. & $333$ ms~(\textcolor{blue}{$153\times$})\\

Jack-o$^\prime$-lanterns (Fig.~\ref{fig:pumpkin}) & $6.7$M & $5.7$M & $4$ & $1/120$ & $5E-4$ & GS & \textcolor{blue}{Penalty} & $32$ & $4$ min. & $753$ ms~(\textcolor{blue}{$40\times$})\\

Squeezed puffer ball (Fig.~\ref{fig:pufferball_net}) & $1.3$M & $0.9$M & $6$ & $1/150$ & $3E-4$ & Jacobi & \textcolor{blue}{Penalty} & $69$ & $67$ min. & $290$ ms~(\textcolor{blue}{$173\times$})\\

Cactus (Fig.~\ref{fig:cactus_dragon}) & $1.2$M & $1$M & $4$ & $1/150$ & $1E-3$ & Jacobi & \textcolor{orange}{IPC} & $40$ & $18$ min. & $171$ ms~(\textcolor{orange}{$82\times$})\\

Animal corossing (Fig.~\ref{fig:animal_crossing}) & $4.8$M & $4.5$M & $4$ & $1/150$ & $1E-3$ & GS & \textcolor{orange}{IPC} & $43$ & $10$ min. & $684$ ms~(\textcolor{orange}{$136\times$})\\

Cloth (Fig.~\ref{fig:cloth}) & $2$M & $3$M & $4$ & $1/120$ & $1E-3$ & Jacobi & \textcolor{orange}{IPC} & $42$ & $48$ min. & $469$ ms~(\textcolor{orange}{$103\times$})\\

\whline{1.15pt}
\end{tabular}
\end{center}
}
\end{table*}

\subsection{Parallel Implementation}
Our method is generic and does not impose restrictions on how a sub-problem should be specified. In our implementation, we assign a sub-problem at each mesh vertex, making $N_i = 3$. The local Newton relaxation needs to tackle a 3-by-3 system, which can be analytically computed. In theory, the positive definiteness of the local Newton system should be taken care of as explained in~\cite{smith2018stable}. We note that the local solve is regularized by $\tilde{\bm{H}}_{i, i}$ i.e., see Eq.~\eqref{eq:reduced_local}, which is sampled at multiple remote Cubature elements. In practice, we would not worry about numerical issues of our local solve since this reduced Hessian is always well-conditioned.

The implementation of Jacobi parallelization is straightforward. When sub-problems are at vertices, one Jacobi iteration solves all the sub-problems at vertices in parallel. No averaging is needed. However, if the sub-problem is defined for multiple vertices e.g., at an element as in~\cite{lan2023second}, a Jacobi iteration also averages duplicated DOFs shared by neighboring sub-problems. Because our local solve is nearly optimal $\delta \bm{x}_i \approx \bm{S}_i \delta \bm{x}^*$, bigger-size sub-problems do not improve the convergence obviously, and lighter local solve should be favored. GS parallelization puts independent sub-problems into groups using graph coloring algorithms~\cite{fratarcangeli2016vivace}. One GS iteration traverses all the sub-problems of all groups. If GS parallelization is used, we consider all the sub-problems within one group as a \emph{generalized sub-problem} and pre-compute the corresponding $\tilde{\phi}_i^k$ for each group as a whole. Because all the sub-problems within a group are independent, the local Hessian is block-diagonal. In other words, the pre-computation effort is as light as the Jacobi parallelization. The only difference lies in the computation of $\bar{\bm{U}}_i$. When GS parallelization is chosen, the constraint in Eq.~\eqref{eq:cms_constraint} applies one unit perturbation at a specific DOF of the current generalized sub-problem, while keeping all the other DOFs of the generalized sub-problem fixed. Nevertheless, we do not observe any noticeable difference between Jacobi and GS parallelization --- both converge at the rate of Newton's method even under large time steps.

\subsection{Overshoot Comparison}

We illustrate the issue of overshoot and compare our method with several well-known parallelable algorithms, including XPBD~\cite{macklin2016xpbd}, projective dynamics (PD)~\cite{bouaziz2014projective}, vertex block descent (VBD)~\cite{chen2024vertex}, and second-order stencil descent (2nd SD)~\cite{lan2023second}. XPBD and PD are widely known for their efficiency and convenient parallelization. A limitation of XPBD or PD is their reliance on the idealization of the material. To make the comparison objective, we use the as-rigid-as-possible (ARAP) material~\cite{igarashi2005rigid}, which can be naturally handled with XPBD and PD.

The direct way to quantify overshoot is to measure the difference between $\bm{x}_i$ and $\bm{S}_i\bm{x}^*$ for sub-problem $i$. We use a standard simulation setup, where a rectangular beam with one end fixed at the wall bends down under its gravity. At a specific frame, we simulate the mesh displacement for the next time step with $h = 1/100$ using global Newton's method. The global convergence condition is set as the relative residual force error being smaller than $1E-4$. The resulting deformation $\bm{x}^*$ serves as the reference for this comparison. After that, we use different algorithms to simulate the same frame with the same material parameters. We pick one tetrahedron element and plot the variation of $\|\bm{x}_i  - \bm{S}_i \bm{x}^*\|$ w.r.t. the number of parallel iterations using different methods. 

\newpage 

\setlength{\columnsep}{5 pt}
\begin{wrapfigure}{r}{0.65\linewidth}
    \includegraphics[width=\linewidth]{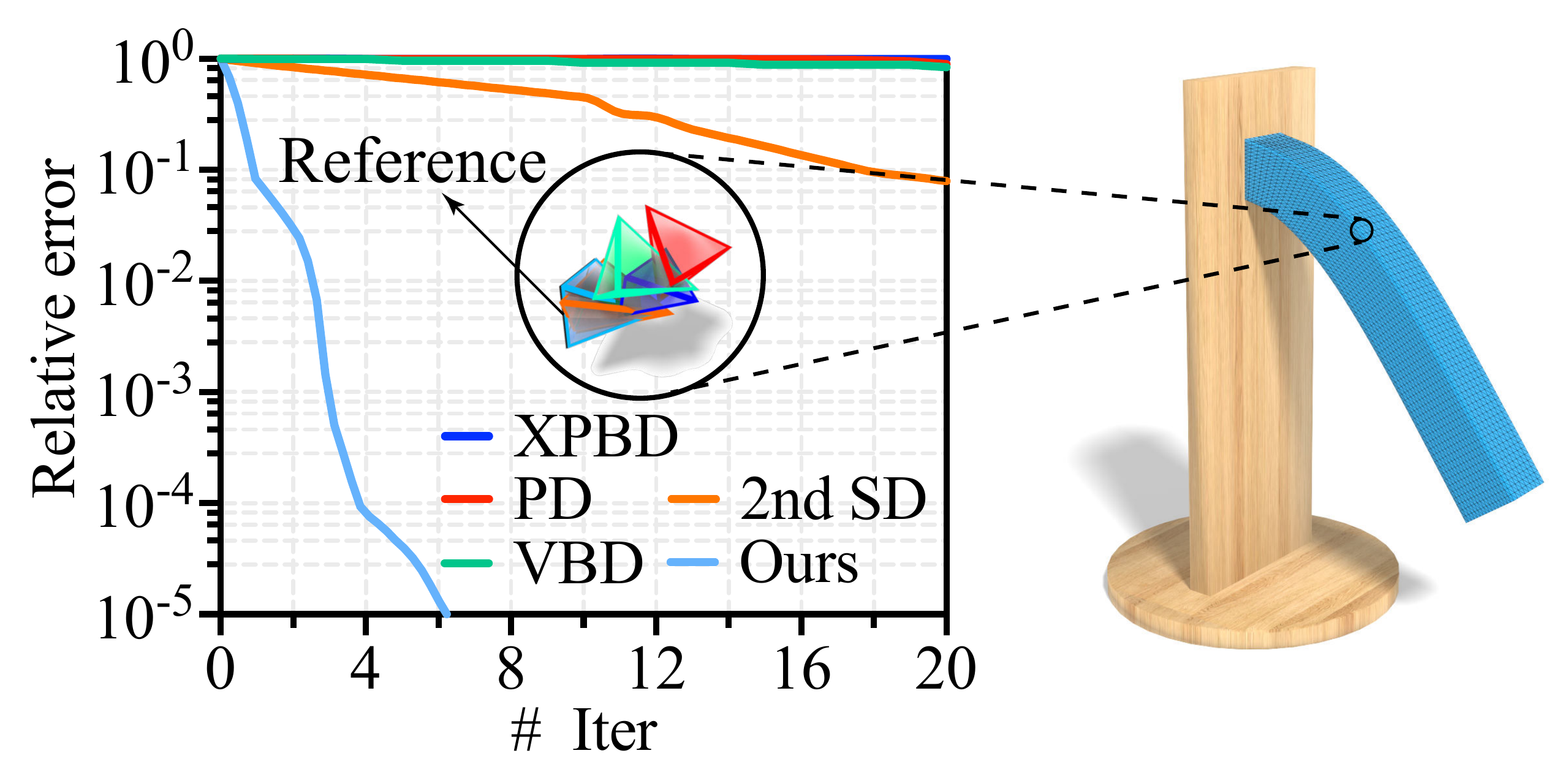}
    \caption{\textbf{Overshoot of local solvers.}~~We plot the relative error between $\bm{x}_i$ and $\bm{S}_i\bm{x}^*$ such that $\bm{x}_i$ is the position of a tetrahedron element obtained by different local solvers, including XPBD~\cite{macklin2016xpbd}, PD~\cite{bouaziz2014projective}, VBD~\cite{chen2024vertex}, 2nd SD~\cite{lan2023second}, and our method. Our method converges as fast as Newton's method does, and the error reaches $1E-3$ with just three iterations. The other methods barely make progress even after 20 iterations.}\label{fig:overshoot}
\end{wrapfigure}

The curves highlight the overshoot problem of existing methods as shown in Fig.~\ref{fig:overshoot}. Our method shows a strong quadratic convergence and pushes $\bm{x}_i$ to the reference with only a handful of iterations. On the other hand, existing methods such as XPBD, PD, and VBD barely improve $\bm{x}_i$ even after 20 iterations. 2nd SD, due to its bigger sub-problem size and hybrid parallelization scheme, delivers a better convergence. But it still gets outperformed by our method by a significant margin. In fact, it will take over $500$ VBD, PD, or XPBD iterations in this example to reduce the relative error to the order of $1E-3$, whereas our method only needs three iterations. The cost of our method for completing one iteration is slightly higher than VBD or XPBD because of extra computations for Cubature elements. Overall, our method is more than $50\times$ faster than XPBD, PD, VBD or 2nd SD in this example. 

\newpage
\begin{wrapfigure}{r}{0.45\linewidth}
    \includegraphics[width=\linewidth]{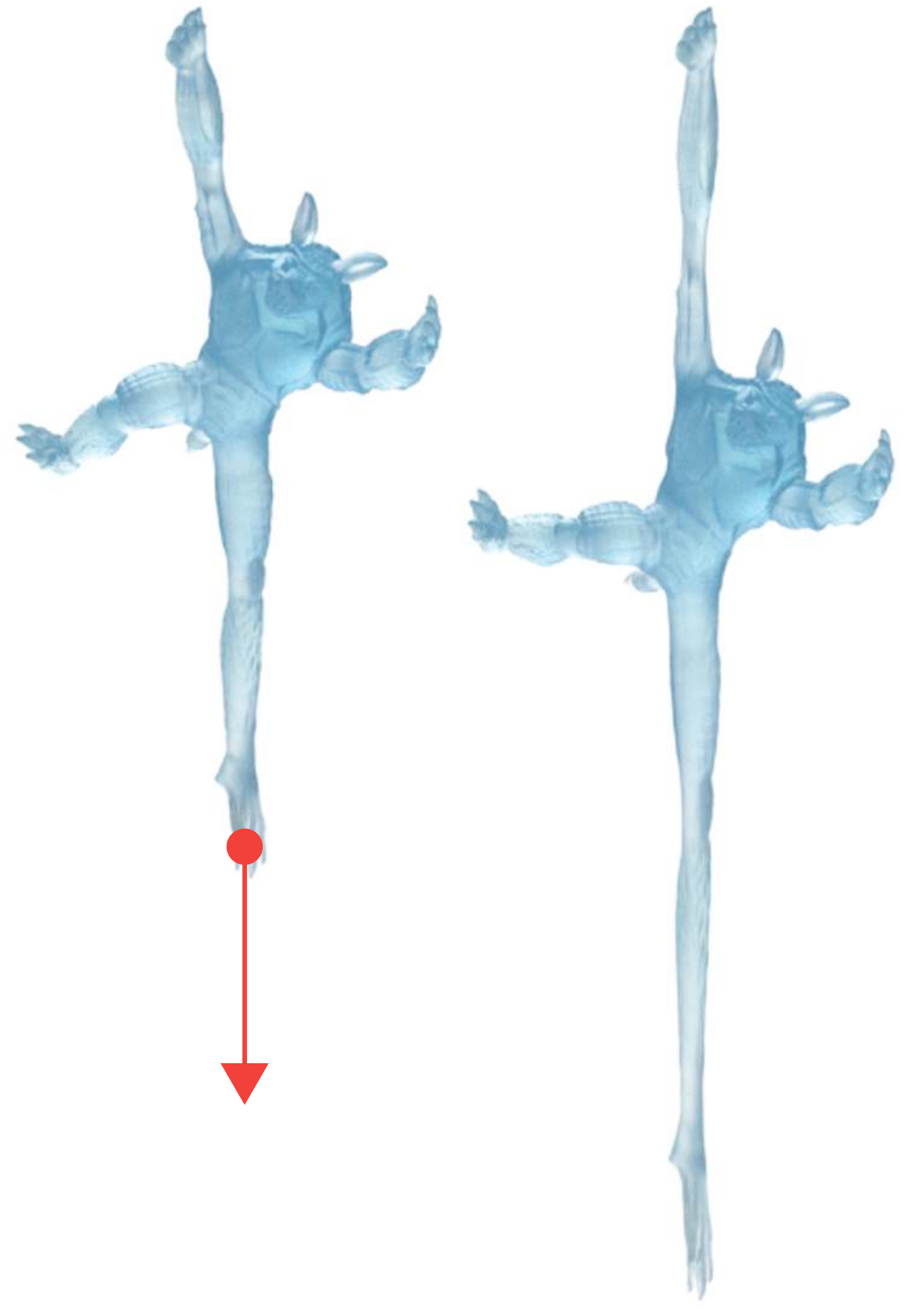}
    \caption{\textbf{Comparison with VBD \& 2nd SD.}~~The Armadillo model consists of $1$M vertices and $3.4$M elements. We fix its left hand and drag the right leg downwards to produce a large-scale body deformation. We record the total number of iterations needed for this example using VBD~\cite{chen2024vertex}, 2nd SD~\cite{lan2022penetration}, and our method under different material stiffness with $h = 1/100$. Our method is $15\times$ faster than 2nd SD and $40\times$ faster than VBD. After further increasing the stiffness of the Armadillo by $20$ times, our method is $34\times$ faster than 2nd SD, and $137\times$ faster than VBD ( VBD switches to a highly conservative time step of $h = 1/20000$).}\label{fig:armadillo}
\end{wrapfigure}

\subsection{Comparison with VBD \& 2nd SD}
We regard VBD from \citet{chen2024vertex} and 2nd SD from \citet{lan2023second} as our most relevant competitors. Both VBD and 2nd SD use Newton's method to handle sub-problems. The key difference is that VBD sets a sub-problem as a vertex (i.e., same as our implementation), while 2nd SD uses a tetrahedron element as a sub-problem. They offer different trade-offs: VBD has better parallelization but converges slower for large time steps or stiff materials. 2nd SD converges better, but local solve is much more expensive, which solves a 12 by 12 system. As shown previously in Fig.~\ref{fig:overshoot}, both VBD and 2nd SD suffer from the overshoot problem. 

To further elaborate on the difference among those peer parallelization strategies, we simulate an Armadillo model with its left hand fixed by applying a sharp and big force at the right foot to generate large-scale body deformation (Fig.~\ref{fig:armadillo}). There are 1M vertices and 3.4M elements on the model. We compare the average number of iterations needed to simulate one time step using VBD, 2nd SD, and our method under different material stiffness with $h = 1/100$. The convergence condition is set as $\|\Delta \bm{x}\| < 1E-4$. The material model is stable Neo-Hookean~\cite{smith2018stable}. On average, it takes $34$ Newton iterations to simulate one time step, and our method needs $38$ (Jacobi) iterations. 
The Newton method takes approximately $150$ seconds per iteration to perform the Cholesky decomposition using the \textsf{MKL PARDISO} solver, while our method takes only $11$ ms to complete an iteration. 2nd SD needs $74$ hybrid iterations. 2nd SD is about $15\times$ more expensive than our method at each iteration. Therefore, our method is $\sim30\times$ faster than 2nd SD. VBD needs $2,264$ iterations to converge one time step, and our method is $40\times$ faster. We then increase the stiffness of the Armadillo for $20$ times and run the same simulation with these methods keeping $h = 1/100$. In this case, the Newton iteration count increases to $58$, and our method needs $64$ iterations. 2nd SD uses $142$ iterations on average, and VBD fails to converge even after $10,000$ iterations. VBD becomes convergent when $h$ is reduced to $1/20000$, and it still needs more than $440$ iterations for one time step. This is equivalent to using $8,800$ iterations to simulate $1/100$ real-world seconds. In this example, our method is $137\times$ faster. The performance of VBD becomes even worse under intensive collisions. 

It now becomes clear that the excellent performance of VBD heavily relied on small time steps and soft materials. 
2nd SD offers a more robust solution for stiff simulations. Unfortunately, both VBD and 2nd SD suffer from overshoot. While our method is orders-of-magnitude faster. The advantage becomes more noticeable in stiffer simulations.

\begin{figure}
    \includegraphics[width=\linewidth]{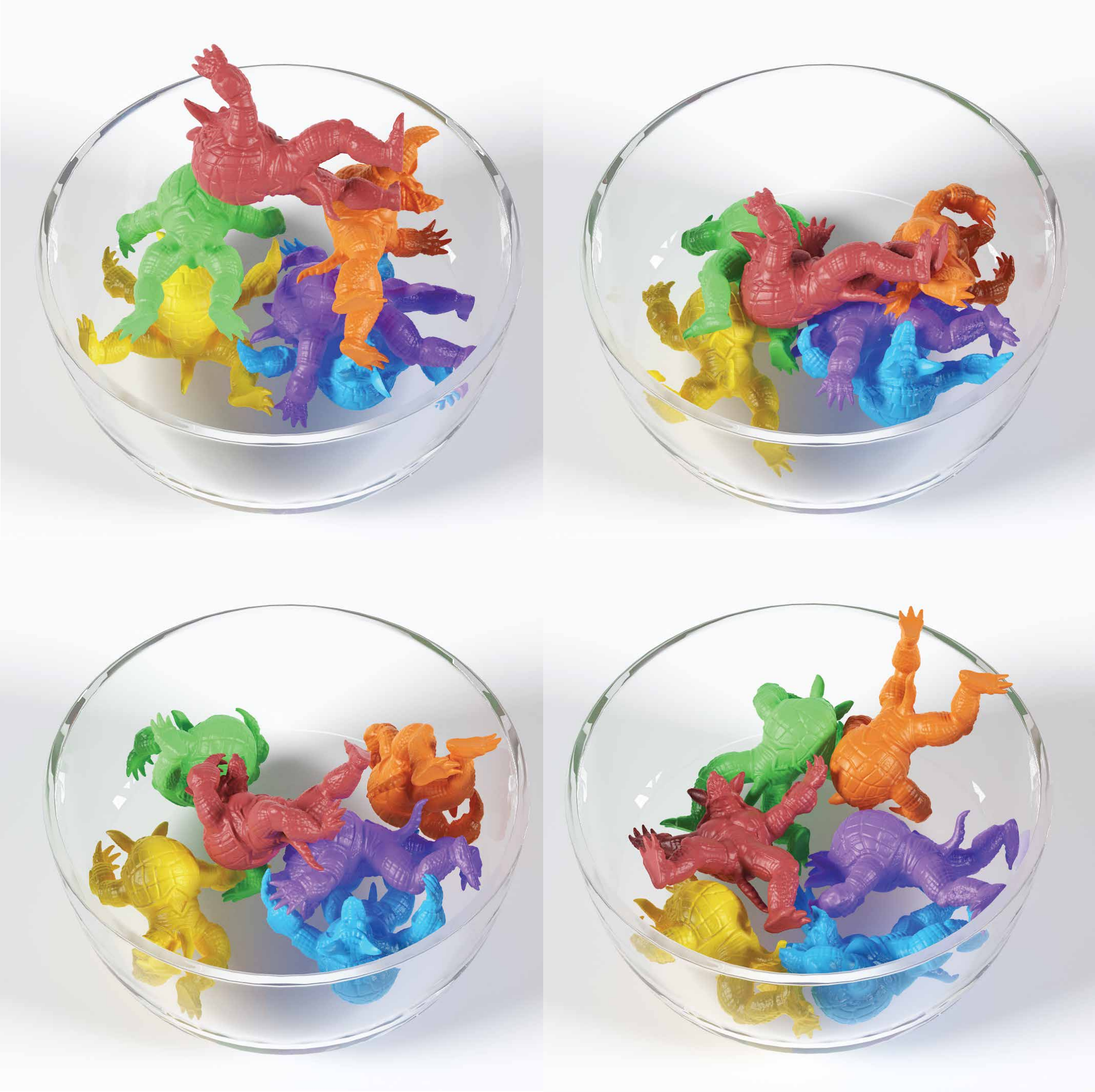}
    \caption{\textbf{Falling Armadillos.}~~Six Armadillo models consist of $1.2$M vertices and $6$M elements. They collide with each other while falling into a container. Our method requires an average of $58$ iterations per time step with $h = 1/150$, compared to $44$ iterations for the projected Newton method. However, our method requires only $883$ ms per step, achieving a speedup of approximately $8000\times$ compared to the projected Newton method, which takes $106$ minutes per step.}\label{fig:dropingAarmadillo}
\end{figure}

\subsection{Comparison with Projected Newton Method}
We also compare our method with the projected Newton method using GPU-based direct solvers. In each iteration, the Newton method performs Cholesky decomposition of a large-scale sparse linear system, which leads to significant computational overhead. In contrast, our method decomposes the global system into smaller $3$ by $3$ subsystems, achieving significant acceleration by fully exploiting GPU parallelism.

As shown in Fig.~\ref{fig:dropingAarmadillo}, we simulate six Armadillo models falling into a container, where Armadillos undergo mutual collisions. These six Armadillos contain $1.2$M vertices and $6$M elements. At this scale, a Cholesky decomposition used for the Newton method takes approximately $150$ seconds. We use $h = 1/150$ and resolve collisions using the implicit penalty method. The projected Newton method requires an average of $44$ iterations to complete a simulation step. Our method requires $58$ iterations under the Jacobi parallelization (e.g., need slightly more iterations compared with Newton). However, each iteration only needs $15$ ms. As a result, our method is approximately $8,000\times$ faster than the projected Newton method.

\begin{figure}
    \includegraphics[width=\linewidth]{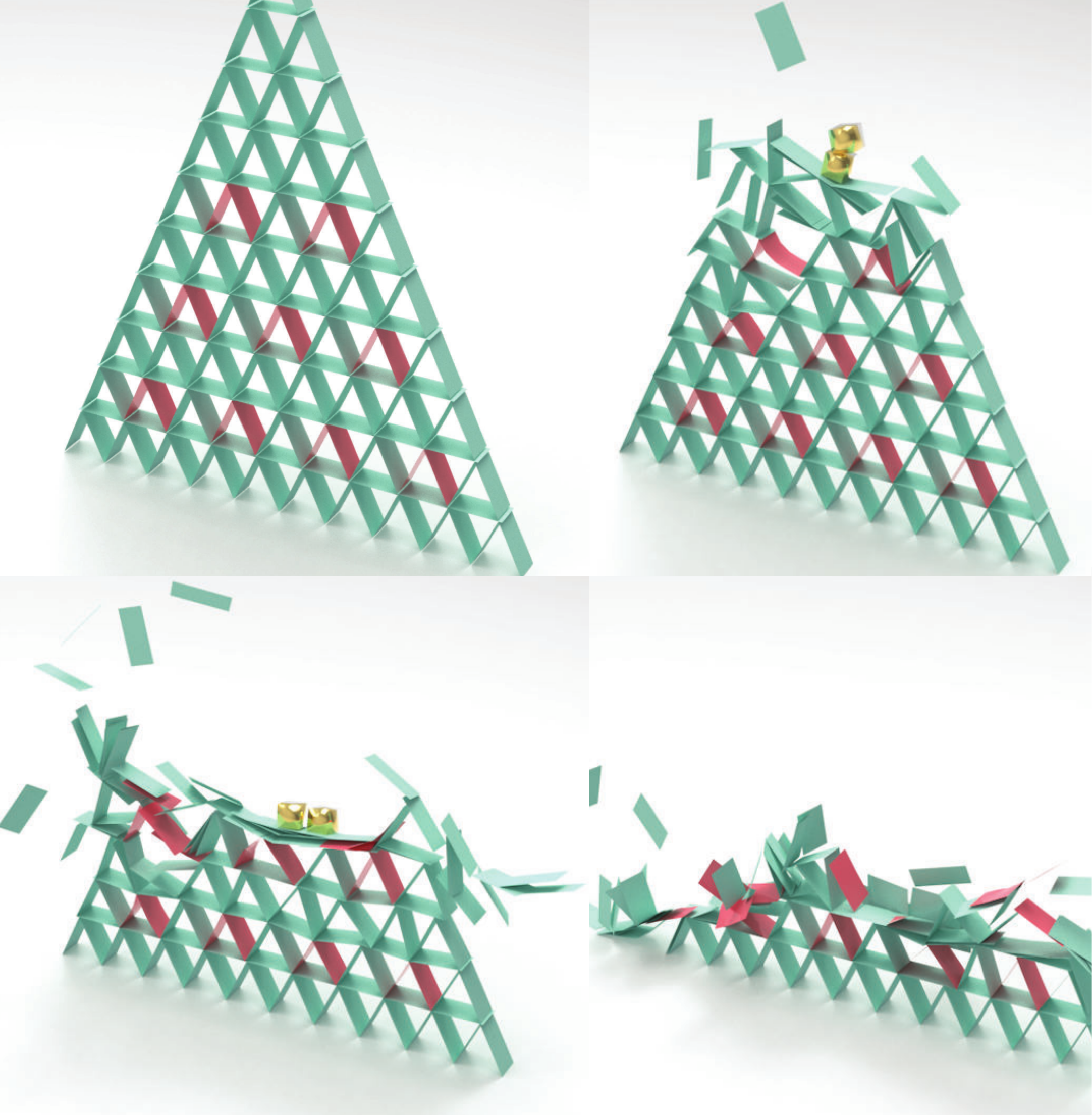}
    \caption{\textbf{House of cards.}~~A stack of $155$ cards is initially balanced through frictional contacts using IPC barriers~\cite{li2020incremental}. The house of cards collapses under a high-velocity impact from two boxes. Each card has $2,543$ elements. The greed cards are $200$ times more stiffer than red ones. Our method takes $31$ ms to simulate one frame, which is over $1,000\times$ faster than CPU-based Newton IPC. VBD does not converge in this example.}\label{fig:cardhouse}
\end{figure}

\begin{figure}
    \includegraphics[width=\linewidth]{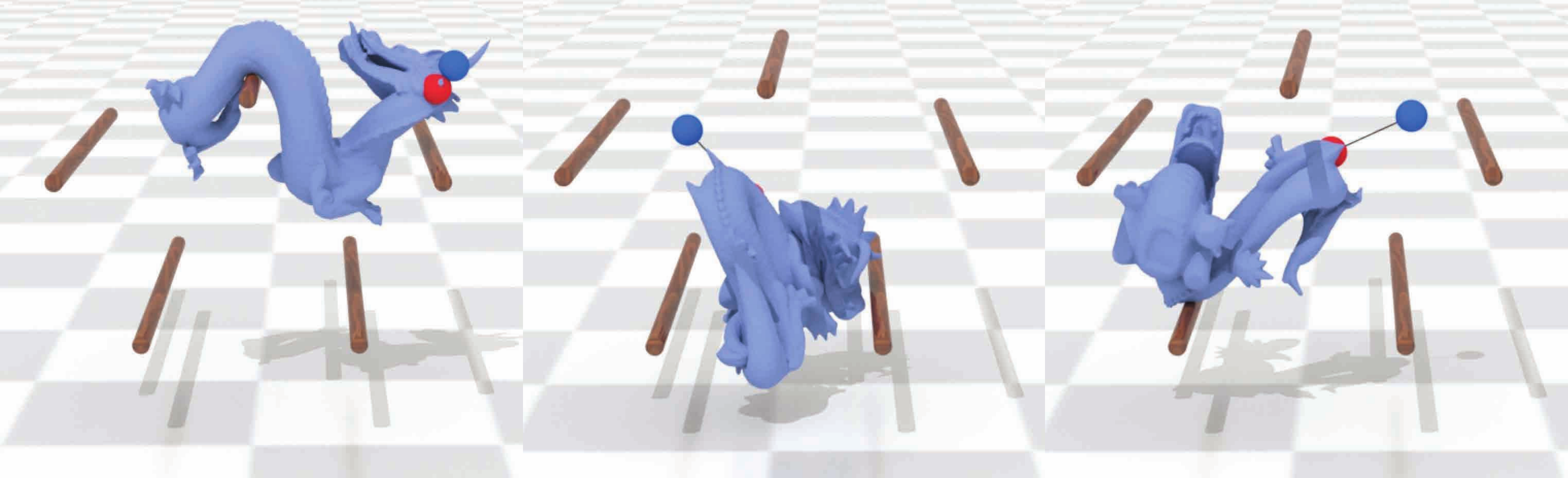}
    \caption{\textbf{Real-time simulation.}~~Our method enables real-time simulation of complex deformable bodies. The dragon has $100$K elements, and it can be simulated in real time under user manipulations. With $h = 1/100$, our method takes fewer than $10$ iterations to simulate one frame. The runtime simulation exceeds $120$ FPS in this example, including collision detection.}\label{fig:realtime}
\end{figure}

\subsection{Collision}
Our method is compatible with existing collision processing algorithms such as IPC~\cite{li2020incremental} or penalty method~\cite{wu2020safe}, as long as the simulation can be formulated as an unconstrained optimization. An example is shown in Fig.~\ref{fig:cardhouse}, where a ``house of card'' is structured by $155$ cards. The green cards are $200$ times more stiffer than red cards. They stack each other the frictional contacts. Two heavy cubes fall, and the card stacking collapses by this external impact. There are $394$K elements in this simulation. Our method uses $31$ ms to simulate one time step ($h = 1/50$), which is over $1,000\times$ faster than CPU-based IPC simulation and more than $120\times$ faster than Newton-Krylov-based GPU solvers~\cite{guo2024barrier}. VBD does not converge in this problem due to the existence of a highly nonlinear barrier (even under $h = 1/5000$). 

\begin{figure}
    \includegraphics[width=\linewidth]{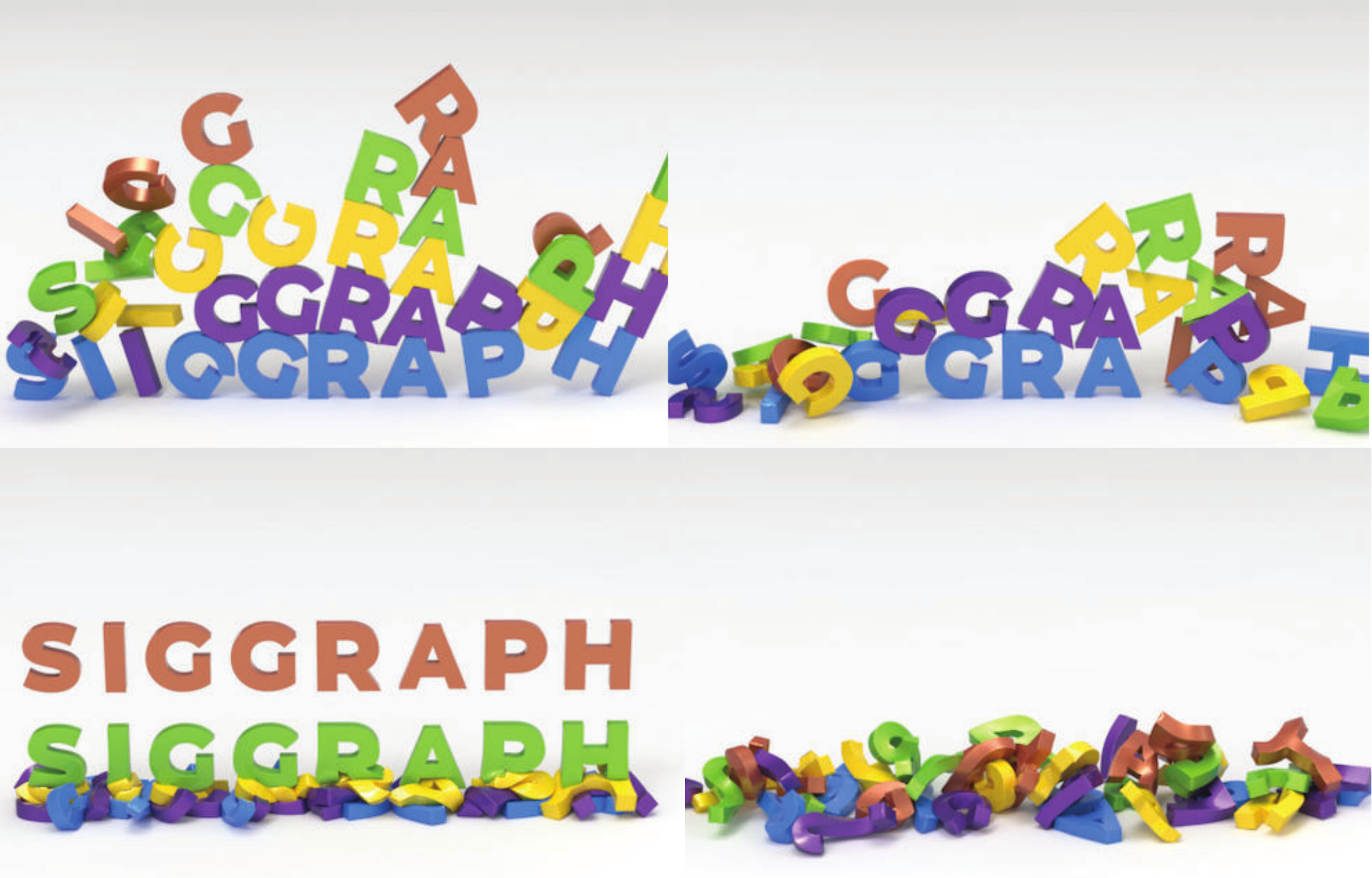}
    \caption{\textbf{Stiff and soft letters.}~~Five sets of ``SIGGRAPH''  letters fall on the floor. The letters on the top row are $1,000$ times more stiffer than the ones in the bottom row. Our method is not sensitive to the variation of material stiffness. For each time step, it needs $34$ iterations for the stiff letters and $27$ iterations for the softer letters. 2nd SD and VBD fail to converge when simulating the stiff letters.}\label{fig:letter}
\end{figure}

\begin{figure*}
    \includegraphics[width=\linewidth]{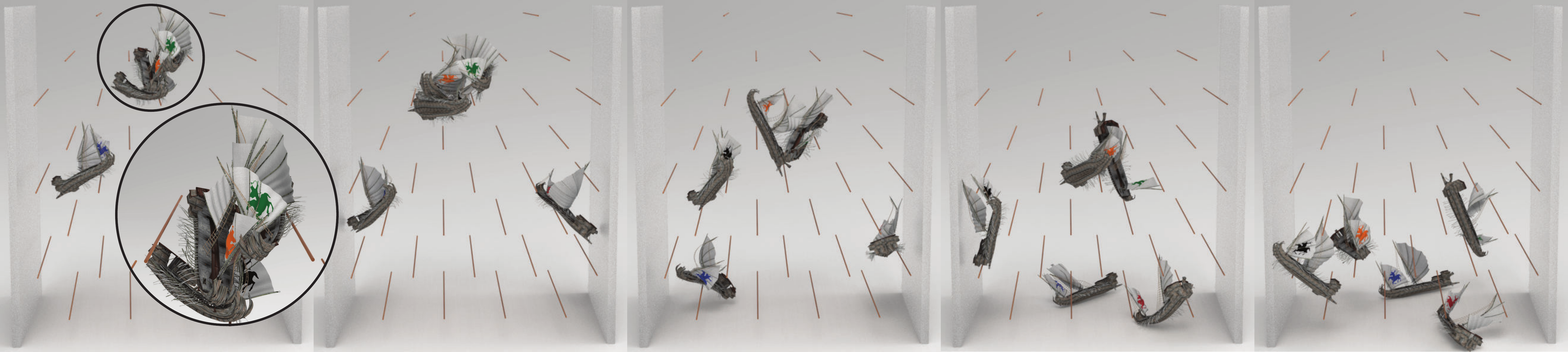}
    \caption{\textbf{Barbarian ships.}~~Five barbarian ships fall and interact with multiple thin rods between two walls. There are more than $2.5$M elements in this example. Our method uses $333$ ms to simulate one time step using penalty forces. Our simulation is $153\times$ faster than VBD.}
    \label{fig:ship}
\end{figure*}

\begin{figure*}
    \includegraphics[width=\linewidth]{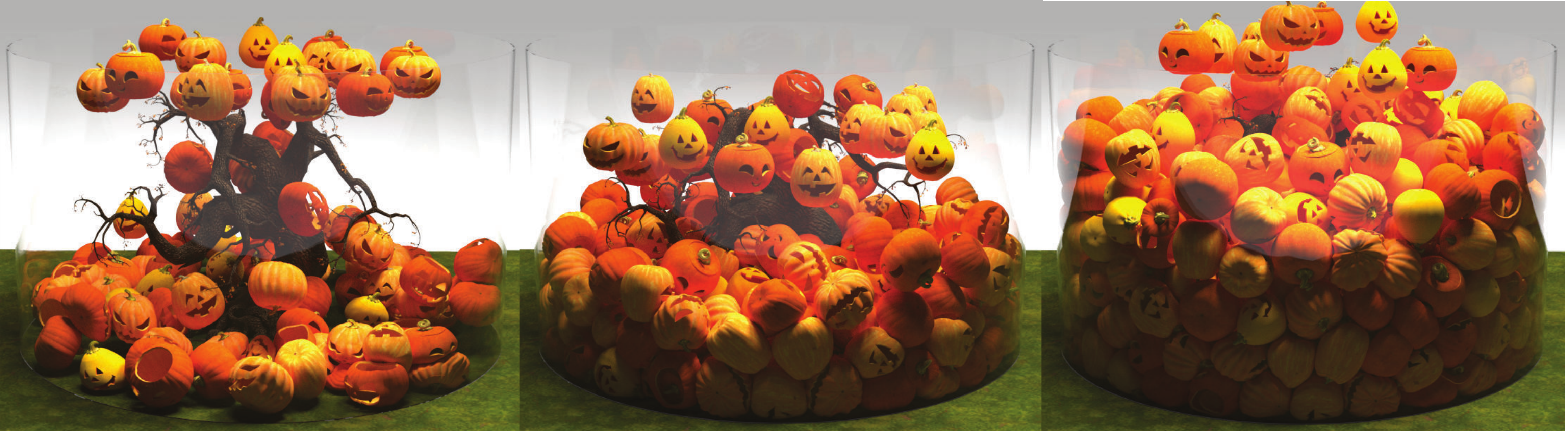}
    \caption{\textbf{Jack-o$^\prime$-lanterns.} $850$ Halloween jack-o$^\prime$-lanterns fall into a container with a deformable tree, and they fully bury the tree eventually. There are $6.7$M elements in this example in total. Our method takes $753$ ms for each time frame, which is $40\times$ faster than VBD.}
    \label{fig:pumpkin}
\end{figure*}

\subsection{Soft \& Stiff Simulations}
Our method exploits $\tilde{\phi}$ to estimate the global energy variation during the local solve. Solving $\bar{\bm{H}}_{C_i, C_i}$ incorporates the material properties of the body at complementary DOFs. This makes our method less sensitive to material variations. Meanwhile, most known parallel algorithms prefer softer simulations over stiffer simulations because DOFs are more strongly coupled with stiff materials, and the local optimum of a sub-problem is more likely to overshoot (e.g., see the comparison of Fig.~\ref{fig:armadillo}). In addition to Fig.~\ref{fig:cardhouse}, we show two more examples in Figs.~\ref{fig:teaser} and \ref{fig:letter} involving both soft and stiff objects. In Fig.~\ref{fig:teaser}, we simulate the dynamics of ten puffer balls sliding from stairs into a glass container from both sides. There are $3.5$M elements in this example. Blue puffer balls are $20$ times softer than the red puffer balls. Our method needs $55$ iterations to simulate one time step ($h = 1/120$). VBD does not converge in this example even under $h = 1/360$. If all the puffer balls are soft ones, VBD becomes converging but is $122\times$ slower than our method. 

Another example is reported in Fig.~\ref{fig:letter}, where we drop five sets of ``SIGGRAPH'' letters on the ground. The letters on the top are $1,000$ times stiffer than the ones at the bottom. Our method handles both simulations stably using similar numbers of iterations i.e., $27$ iterations for soft letters and $34$ iterations for hard letters). Meanwhile, neither VBD nor 2nd SD converges for hard letters.

\subsection{Real-time Simulation}
Our method enables real-time elastic simulation of complicated shapes of real-world material. Fig.~\ref{fig:realtime} shows snapshots of screen records of a real-time simulation of a dragon. The user interactively manipulates the Neo-Hookean dragon~\cite{smith2018stable}, which consists of $100$K elements, and the simulation runs in real-time at more than $100$ FPS following the user inputs.

\begin{figure*}
    \centering
    \includegraphics[width=\linewidth]{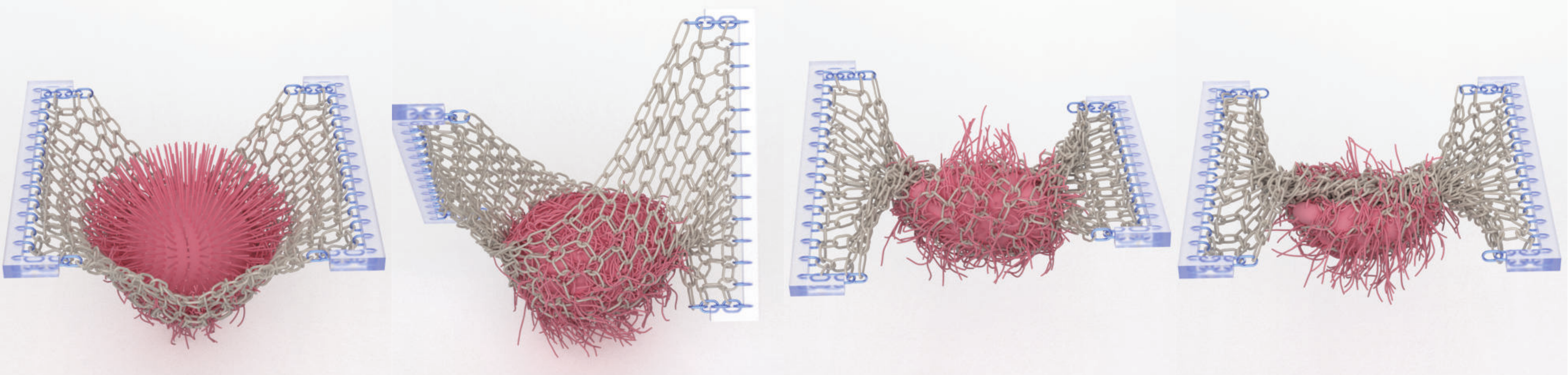}
    \caption{\textbf{Squeeze puffer ball.}~~In this simulation, we drop a puffer ball into a soft elastic chain. The puffer ball, comprising $1.2$M elements, interacts with an elastic net made up of $588$ rings and $329$K elements, connected via ring-ring contacts. As the ball descends, the net twists and compresses it, demonstrating the effects of tightly coupled contacts and elasticity. This simulation runs with a time step of $h = 1/150$, and each time step takes $290$ ms. Our method is $173\times$ faster than VBD.}
    \label{fig:pufferball_net}
\end{figure*}

\begin{figure*}
    \centering
    \includegraphics[width=\linewidth]{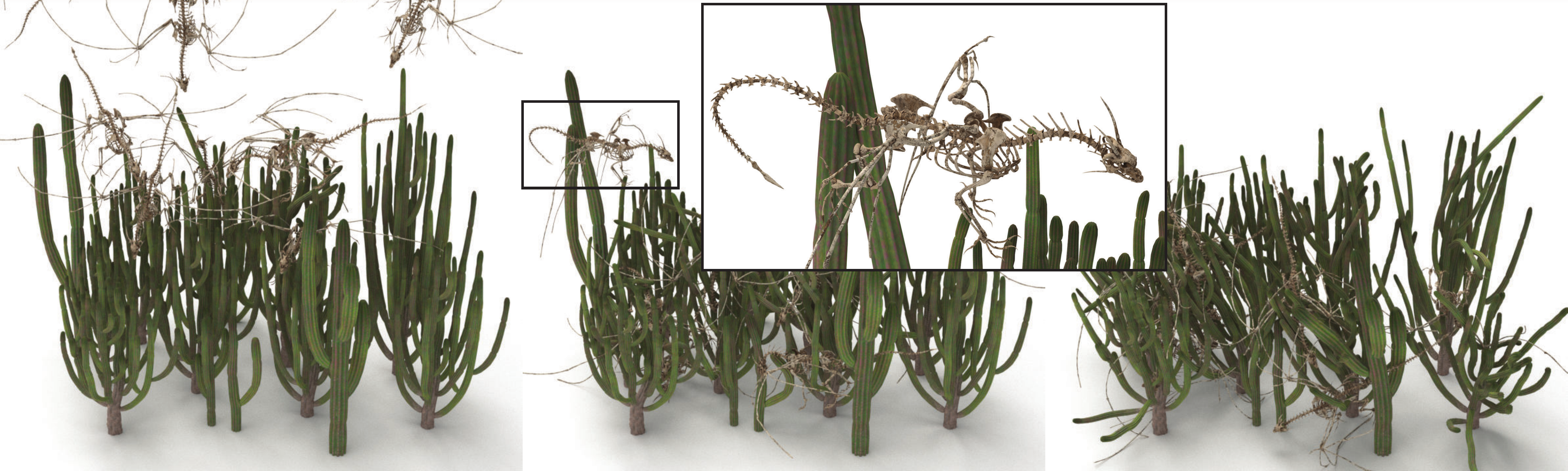}
    \caption{\textbf{Bone dragons from the sky.}~~Six bone dragons fall into the cactus bush. There are $1.2$M elements in the simulation, and we use IPC to robustly process high-velocity collisions between bone dragons and cacti, as well as self-collisions among cacti. In this example, our method is over $82\times$ faster than GPU-based IPC simulation~\cite{guo2024barrier}.}
    \label{fig:cactus_dragon}
\end{figure*}

\subsection{More Results}
We show more large-scale simulation results using our method. All of these examples involve complex shapes and high-resolution models. Fig.~\ref{fig:ship} reports an example of simulating five barbarian ships. Each barbarian ship has $500$K elements, and the simulation handles $2.5$M elements. Under $h =  1/120$, our method needs $34$ iterations on average, and the solving time for each time step is $333$ ms. This is $153\times$ faster than VBD. In Fig.~\ref{fig:pumpkin}, we keep dropping deformable Jack-o$^\prime$-lanterns into a container with an old tree until the tree is fully buried by the lanterns. There are $6.7$M DOFs and $850$ pumpkins in this simulation. Our method is $40\times$ faster than VBD. 

The reader may notice the different performance gains in those two examples, and the improvement of our method tends to become less pronounced in Fig.~\ref{fig:pumpkin}. This is because deformable bodies in Fig.~\ref{fig:pumpkin} do not have a large number of elements compared with the ship model in Fig.~\ref{fig:ship}. Isolated DOFs on different objects are naturally decoupled, and the local solver is less likely to overshoot. To verify this, we show another challenging simulation in Fig.~\ref{fig:pufferball_net}. In this example, we simulate a puffer ball with $1.2$M elements (i.e., it is of higher resolution than the puffer ball model used in Fig.~\ref{fig:teaser}) falling into an elastic net of $588$ rings and $329$K elements. The net twists and squeezes the puffer ball. In this example, our method is $173\times$ faster than VBD. 

\begin{figure*}
    \centering
    \includegraphics[width=\linewidth]{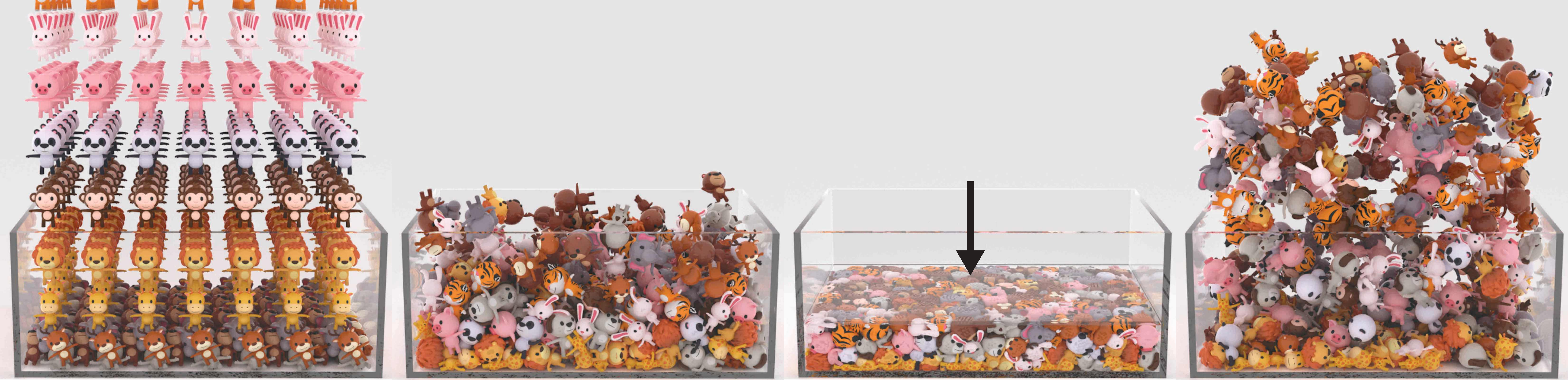}
    \caption{\textbf{Animal crossing.}~~500 small animal toys fall into the tank, and there are $4.8$ elements in this example. A glass plane is then pushed down to compress all these little toys. After the removal of the plane, the compressed animals bounce back into the air. Animals have different stiffness Our method takes $684$ ms to simulate one time step, which is $70\times$ faster than GPU projective dynamics~\cite{lan2022penetration}, and $136\times$ faster than GPU-IPC~\cite{guo2024barrier}.}
    \label{fig:animal_crossing}
\end{figure*}

Our method works with IPC as well~\cite{li2020incremental}. An example is given in Fig.\ref{fig:cactus_dragon}, where six bone dragons fall into a cactus bush with high velocities. The impacts from the dragon trigger significant dynamics at the cactus. The use of IPC ensures the simulation is free of interpenetration, and frictional contacts between the bone dragons and cactus are also accurately captured. Compared with vanilla IPC, which solves the global Newton at each iteration, our method is about $5,000\times$ faster. This speedup is a rough estimation, as we have not completed this simulation on CPU IPC. Compared with GPU-IPC~\cite{guo2024barrier}, our method is $82\times$ faster. There are $1$M DOFs in the simulation, and our method takes $171$ ms to simulate one frame. Another example is shown in Fig.~\ref{fig:animal_crossing}, where $500$ small animals drop into a tank. After that, we use a glassy plane to press all the animals and release the constraint suddenly, making all the animal toys bounce back up. Those little toys are of different stiffness. In this example, there are $4.8$M elements, and it is $70\times$ faster than GPU projective dynamics~\cite{lan2022penetration}.

\begin{figure*}
    \centering
    \includegraphics[width=0.8\linewidth]{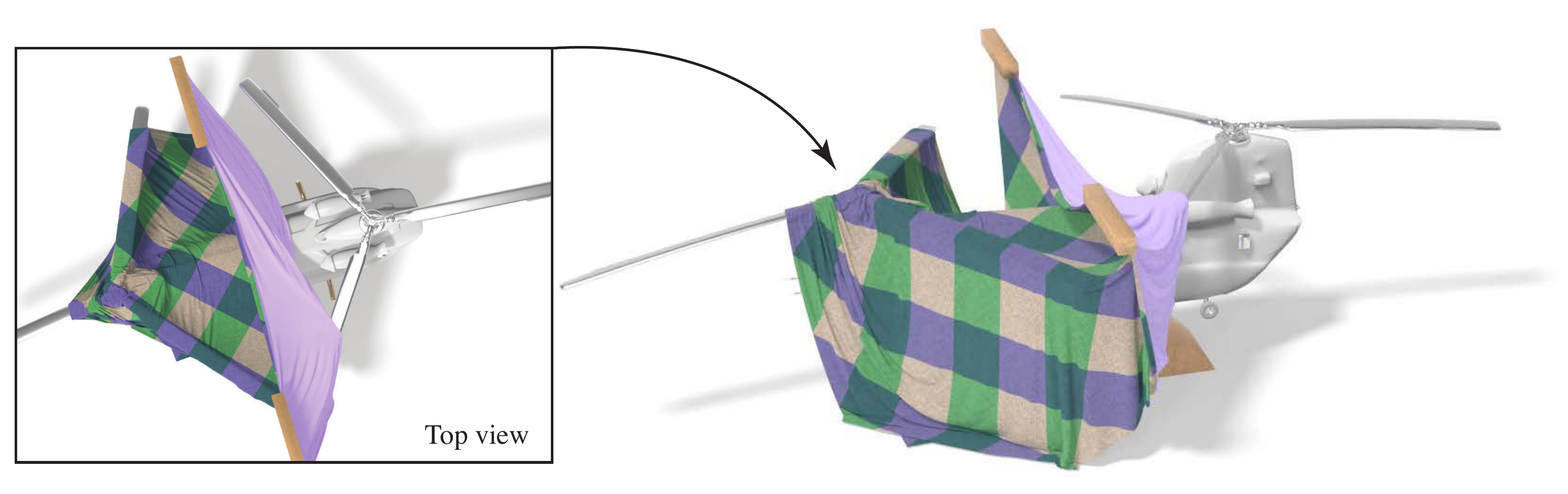}
    \caption{\textbf{Cover the helicopter.}~~Our method is not limited to deformable simulation and can also be readily used for thin-shell and cloth simulation. In this example, a piece of tablecloth covers a helicopter. It is displaced back and forth, generating detailed wrinkles. There are $3$M DOFs in the simulation. We use IPC to process collisions between the cloth and the helicopter as well as the self-collision on the cloth. It takes $469$ ms to simulate one frame, and our method is $12,00\times$ faster than co-dimensional IPC~\cite{li2020codimensional}.}
    \label{fig:cloth}
\end{figure*}

Our method is also able to handle co-dimensional models like thin-shell and cloth. To this end, we show the result of a high-resolution cloth simulation in Fig.~\ref{fig:cloth}. The tablecloth consists of $2$M triangles and over $3$M DOFs. It is displaced to cover a helicopter model from the top. Our method captures detailed wrinkles of the cloth during the movement, and it converges with about $42$ iterations on average. In this example, our method is three orders faster than CPU-based C-IPC~\cite{li2020codimensional}.

%% file: conclusion.tex
\section{Conclusion \& Future Work}\label{sec:conclusion}
In this paper, we explore the problem of overshoot in parallel deformable simulation. When overshoot occurs, the local solver over-aggressively reduces its local target, and the resulting DOFs' update negatively impacts the target function at other areas on the deformable body. We give a second-order optimal solution to avoid overshoot and make this procedure pre-computable. This leads to a new GPU simulation algorithm that possesses both excellent parallelism and (near) second-order convergence. We have tested our method on a wide range of large-scale simulation scenes. Our method constantly outperforms existing GPU simulation algorithms by orders, making real-time simulation of complicated deformable objects possible.

Our method also has some limitations. Our method needs to build a local subspace at each sub-problem and carry out Cubature training for co-rotationed basis vectors. Such pre-computation is slow and could impose practical inconvenience. The second-order convergence depends on the quadratic approximation of the global optimal $E^\star$ i.e., see Eq.~\eqref{eq:newton}. When Newton's approximation is less appropriate and $\|\delta \bm{x}^k\|^3$ is a relatively big quantity, our method does not converge quadratically. This could happen when the variational optimization involves highly nonlinear terms such as the IPC barrier. Line search is then needed. Fortunately, line search can be done at each sub-problem in parallel. As a result, our method remains orders-of-magnitude faster than global IPC solvers. While our algorithm pushes the simulation performance to a new level, collision detection becomes the new bottleneck. We demonstrate the feasibility and potential of our method in the context of hyperelastic simulation, yet the proposed algorithm is actually a general-purpose parallel optimization procedure. Therefore, it is of great interest for us to apply our method in other simulation and graphics problems. The key difficulty is still how to find a pre-computable $\phi$ to efficiently and effectively estimate the global energy variation. We believe a data-driven approach, i.e., a deep learning perspective may be a good answer to this challenge, which could offer a case-by-case optimized setup for different computational problems.

\begin{acks}
We thank reviewers for their detailed and constructive comments. Chenfanfu Jiang is partially supported by NSF 2153851 and TRI. Yin Yang is partially supported by NSF under grant number 2301040. 
\end{acks}